\documentclass[aps,twocolumn,superscriptaddress,showpacs,floatfix]{revtex4}
\usepackage[latin9]{inputenc}
\usepackage{amsmath}
\usepackage{amssymb}
\usepackage{graphicx}
\usepackage{esint}
\usepackage{epstopdf}
\usepackage{hyperref}

\makeatletter
\@ifundefined{textcolor}{}
{
 \definecolor{BLACK}{gray}{0}
 \definecolor{WHITE}{gray}{1}
 \definecolor{RED}{rgb}{1,0,0}
 \definecolor{GREEN}{rgb}{0,1,0}
 \definecolor{BLUE}{rgb}{0,0,1}
 \definecolor{CYAN}{cmyk}{1,0,0,0}
 \definecolor{MAGENTA}{cmyk}{0,1,0,0}
 \definecolor{YELLOW}{cmyk}{0,0,1,0}
}

\makeatother

\begin{document}

\title{Birth of a quasi-stationary black hole in an outcoupled Bose-Einstein condensate}

\author{J. R. M. de Nova}

\affiliation{Departamento de Física de Materiales, Universidad Complutense de
Madrid, E-28040 Madrid, Spain}
\thanks{jrmnova@fis.ucm.es}
%\affiliation{Departamento de Física de la Materia Condensada, Universidad Aut\'onoma de
%Madrid, E-28049 Madrid, Spain}

\author{D. Gu\'ery-Odelin}

\affiliation{
Universit\'e de Toulouse, UPS, Laboratoire Collisions Agr\'egats R\'eactivit\'e, IRSAMC, and CNRS, UMR 5589,
F-31062 Toulouse, France
}

\author{F. Sols}

\affiliation{Departamento de Física de Materiales, Universidad Complutense de
Madrid, E-28040 Madrid, Spain}

\author{I. Zapata}

\affiliation{Departamento de Física de Materiales, Universidad Complutense de
Madrid, E-28040 Madrid, Spain}

\begin{abstract}

We study the evolution of an initially confined atom condensate which is progressively outcoupled by gradually lowering the confining barrier on one side. The goal is to identify protocols that best lead to a quasi-stationary sonic black hole separating regions of subsonic and supersonic flow. An optical lattice is found to be more efficient than a single barrier in yielding a long-time stationary flow. This is best achieved if the final conduction band is broad and its minimum not much lower than the initial chemical potential. An optical lattice with a realistic Gaussian envelope yields similar results. We analytically prove and numerically check that, within a spatially coarse-grained description, the sonic horizon is bound to lie right at the envelope maximum.
We derive an analytical formula for the Hawking temperature in that setup.

\end{abstract}


\date{\today}

\maketitle

\section{Introduction}

An attractive feature of Bose-Einstein condensates is that of providing a convenient way of investigating analog black-hole physics in the laboratory \cite{Garay2000,Chapline2003}.
%((add Chapline G, Laughlin R B and Santiago D I 2003 Analog Models of General Relativity ed M Visser (Singapore World Scientific) p. 179 \cite{Chapline2003}
%%179-98
%)).
It was already noted by Unruh \cite{Unruh1976,Unruh1981} that Hawking radiation in a cosmic black hole \cite{Hawking1974,Hawking1975} is an essentially kinematic effect that could be simulated in a quantum fluid. More specifically, it has been predicted that,
for a quantum fluid passing through a sonic horizon (i.e., a subsonic-supersonic interface), phonons will be emitted into the subsonic region even at zero temperature
\cite{Leonhardt2003a,Leonhardt2003,Balbinot2008,Carusotto2008,Macher2009a,Finazzi2010}.

%A common feature of these proposals is that they rely on the achievement of a stationary sonic black hole at the interface between subsonic and supersonic flow (\cite{Leboeuf2001}). A canonical scenario for this type of flow is that of a leaking condensate. The subsonic flow occurs in the region where the condensate is initially confined, while the supersonic region is identified with the low density gas that flows away from the main region.
A sonic black hole has been recently realized in an accelerated
Bose-Einstein condensate \cite{Lahav2010}.
An alternative route towards the detection of Hawking radiation
may be provided by a quasi-stationary horizon, which in principle can be
achieved by allowing a large confined condensate to emit in
such a way that the coherent outgoing beam is dilute and fast enough to
be supersonic \cite{Recati2009,Macher2009a,Zapata2011,Larre2012}.
The hope is that the so far elusive Hawking radiation will be less difficult to detect unambiguously
in such quasi-stationary transport scenarios.
In particular, some recent works have addressed the possibility of detecting the spontaneous Hawking radiation above the stimulated signal \cite{Finazzi2013,deNova2014,Busch2014}.
Experimental evidence of stimulated Hawking radiation has been recently found in a black-hole laser setup \cite{Steinhauer2014}.

The main goal of this paper is to explore the actual attainability of the steady-state regime. Within a mean-field approximation, we investigate the dynamics of an initially confined condensate that begins to leak as the height of one of the confining barriers is driven from an essentially infinite to a finite value that permits a gentle yet appreciable flow of coherently outcoupled atoms. A similar scenario has been already considered in Ref. \onlinecite{Wuster2007}. An alternative route to a quasi-stationary black-hole configuration has been proposed for atom \cite{Kamchatnov2012} and polariton \cite{Gerace2012} condensates, based on the idea of throwing the condensate onto a localized obstacle such as a potential barrier. In the present work, we focus on a finite-sized condensate and on the case where the increasingly transparent potential is formed not by a single \cite{Larre2012} or double \cite{Zapata2011} barrier, but by an extended optical lattice, the main reason being that the latter scenario seems more suitable for the achievement of quasi-stationary flow within this deconfinement scheme, as will be shown later.  We will see that close-to-ideal stationary flow within the permitted energy bands is achievable under realistic opening protocols.

The present mean-field study aims at identifying transport scenarios that offer hopes for a future detection of Hawking radiation. More conclusive predictions about the detectability of spontaneous radiation will require a study of the time-dependent Bogoliubov - de Gennes (BdG) equations, which should inform us on the actual intensity of the expected spontaneous radiation. This task is left for a future study.
%Thus, in the present work, when we qualify functions of position and time as e.g. slow, flat, or constant, we follow internal criteria defined within this paper but do not mean that they behave well enough to allow for the detection of Hawking radiation.

%Here, lacking the results of that work and across the present work, we will leave unqualified many quantities in statements such as \emph{sufficiently slow, flat, constant, ... } i.e., without comparing to other physical quantities. Computed quasi-particle characteristics (those necessary to make up Hawking radiation spectrum) will offer the relevant time and space variations to which these unqualified quantities should be matched.

Besides the motivation of realizing gravitational analogs, the achievement of stationary transport scenarios is of general interest for the investigation of atom quantum transport, in the case of both bosons \cite{Bloch1999,Andersson2002,PhysRevA.76.063605,Guerin2006,PhysRevA.80.041605,PhysRevA.84.043618} and fermions \cite{Brantut2012,Brantut2013}, within the emergent field of atomtronics.

This paper is arranged as follows. Section \ref{sec:themodel} presents the model for the gradual reduction of the optical lattice amplitude which we will be investigating. After some preliminary remarks in Section \ref{sec:Prerem}, the main numerical results together with some theoretical arguments (that help to understand the observed trends) are presented in Section \ref{sec:numerical}. The second part of that section describes the achieved quasi-stationary regime. Section \ref{sec:Gaussian-shaped} addresses the more realistic case of an optical lattice having a Gaussian envelope. Interestingly, we find that the horizon lies at the maximum of the envelope and give a theoretical explanation of that fact.
The main conclusions are summarized in section \ref{sec:conclusions}. Appendix \ref{app:GPstationary} provides a detailed description of the initial state of the condensate as it exists before the deconfinement procedure begins. Appendix \ref{app:nonlinearol} discusses some properties of Bloch waves in the presence of nonlinearities accounting for the interaction. Appendix \ref{app:olperturbationtheory} presents a perturbative treatment of the interaction. Finally, Appendix \ref{app:crnicabc} describes the numerical method of integration and the use of absorbing boundary conditions.

\section{The model} \label{sec:themodel}

In this work we study the outcoupling of a one-dimensional (1D) Bose-Einstein condensate
through a finite-size repulsive optical lattice, whose intensity is gradually lowered in such a way that, within a finite time, the periodic barrier shifts from
a regime of practical confinement to one of full transparency within certain atom energy bands.
We focus on the mean-field dynamics, i.e., we only consider the evolution of the condensate wave function, leaving the dynamics of quasi-particles
for a future study. We restrict our present study to a quasi-one dimensional
model. The time-dependent condensate wave function $\Psi(x,t)$
obeys the Gross-Pitaevskii (GP) equation \cite{Pethick2008,Pitaevskii2003}:
%\begin{widetext}
%\begin{eqnarray}
%i\hbar\frac{\partial\Psi(x,t)}{\partial t} & = & \left[-\frac{\hbar^{2}}{2m}\partial_{x}^{2}+V(x,t)+g|\Psi(x,t)|^{2}\right]\Psi(x,t)\, ,\label{eq:TDGP}
%\end{eqnarray}
%\end{widetext}
\begin{eqnarray}
& & i\hbar\frac{\partial\Psi(x,t)}{\partial t} \nonumber \\
& = & \left[-\frac{\hbar^{2}}{2m}\partial_{x}^{2}+V(x,t)+g|\Psi(x,t)|^{2}\right]\Psi(x,t)\, ,\label{eq:TDGP}
\end{eqnarray}
where $m$ is the atom mass and $V(x,t)$ the
time-dependent optical lattice potential. The effective one-dimensional coupling constant $g=2\hbar\omega_{\rm tr}a_{s}$ (with $a_{s}$ the $s$-wave scattering
length) is the relevant interaction strength in a setup where only the ground state of a confining transverse harmonic oscillator of frequency
$\omega_{\rm tr}/2\pi$ is populated. This is the 1D mean-field regime, characterized by the condition $\rho a_s\ll 1$, where $\rho(x,t)=|\Psi(x,t)|^2$ \cite{Menotti2002,Leboeuf2001}.
At the same time, $\rho a^2_{\rm tr}/a_s\gg 1$ (with $a_{\rm tr}$ the transverse oscillator length) must be satisfied to stay away from the Tonks-Girardeau regime \cite{Menotti2002,Petrov2000,Dunjko2001}. Taking the initial bulk density
$n_0$ as a typical value for the density, we can realistically set (see Sec. \ref{sec:numerical}) $n_0a_s\sim10^{-1}$ and $n_0 a^2_{\rm tr}/a_s\sim10^{3}$, from which we conclude that we are safely in the 1D mean-field regime.

Equation (\ref{eq:TDGP}) conserves the total particle number $N$ as given by the normalization condition
\begin{equation}
N=\int\mathrm{d}x|\Psi(x,t)|^{2} \, . \label{eq:normalization}
\end{equation}
The condensate density is nonzero only for $x>0$
because at all times we assume a sufficiently high barrier at $x=0$, which is simply
implemented via the hard-wall boundary condition $\Psi(0,t)=0$.

Initially (at times $t < 0$), we consider an equilibrium condensate made of $N$
atoms occupying the region $0<x \alt L$. Thus $n_{0}\simeq N/L$ is the initial atom density, which is defined below more precisely. We also
introduce an optical lattice that spans the region $L \alt x \alt L+L_{\rm lat}$ and whose initial amplitude $V_0$ is
large enough for particle tunneling through the lattice to be practically forbidden.
The initial wave function is stationary, $\Psi(x,t)=e^{-i\mu_{0}t/\hbar}\Psi(x)$,
with $\Psi(x)$ satisfying the time-independent GP
equation
\begin{eqnarray}
\left[-\frac{\hbar^{2}}{2m}\partial_{x}^{2}-\mu_{0}+V(x,0)+g|\Psi(x)|^{2}\right]\Psi(x) & = & 0 \, .\label{eq:GP}
\end{eqnarray}
The initial chemical potential $\mu_0$ is determined by the normalization condition (\ref{eq:normalization}).
The initial healing length is defined as $\xi_0\equiv \sqrt{\hbar^{2}/mgn_{0}}$, where $n_{0}\equiv \mu_{0}/g$. Further details on the initial condensate are given in Appendix \ref{app:GPstationary}. At time $t=0$, the optical lattice intensity starts to decrease and atoms begin to escape towards the region $x \agt L+L_{\rm lat}$, where the potential is assumed to be negligible.
On quite general grounds \cite{Leboeuf2001,Zapata2009a,Zapata2011}, the
flow beyond the optical lattice can be expected to be supersonic.

We assume that the optical lattice is
made of two fixed phase
lasers of wavelength $\lambda$ and whose wave vectors form an angle $\theta$ \cite{Fabre2011,Blakie2002}.
The time-dependent optical lattice potential is chosen so that in the
lattice region (defined by $L-\frac{d}{2}\leq x\leq L-\frac{d}{2}+L_{\rm lat}$) and
for times $t\geq0$,
\begin{eqnarray}
V(x,t) & = & V(t)\cos^{2}\left[k_L(x-L)\right]\nonumber \\
V(t) & = & V_{\infty}+(V_{0}-V_{\infty})e^{-t/\tau} \, ,\label{eq:TDPotential}
\end{eqnarray}
where $k_L=\pi/d$ and $d=\lambda/\left[2\sin(\theta/2)\right]$ is the lattice period, while $V(x,t)=0$ everywhere else.

\begin{figure}[tb!]
\includegraphics[width=1\columnwidth]{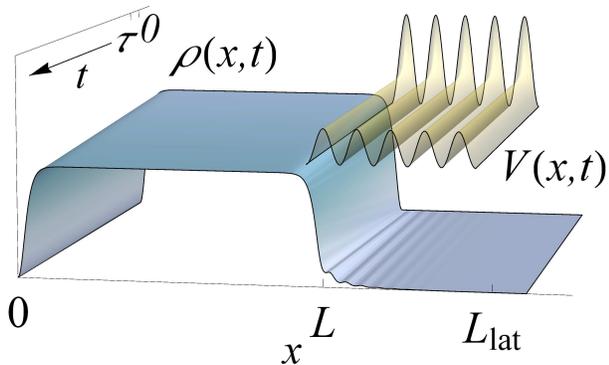}
\caption{Schematic representation of the emitting condensate setup studied in this article.
Within the ideal lattice scenario, hard-wall boundary conditions are assumed at $x=0$ and
an optical lattice lies in the region $L<x<L_{\rm lat}$ with a time-dependent amplitude such that the potential
$V(x,t)$ (represented by the semi-transparent yellow surface over the $x-t$ plane) evolves from strongly to moderately confining. The resulting time-dependent density profile $n(x,t)$ is represented by the grey-blue surface.
The vertical axis gives the density in some (here unimportant) units. The surface $V(x,t)$ is uplifted to provide a better vision of $n(x,t)$. Some parameters defined in the main text are indicated.
The trend towards a long-time quasi-stationary flow regime can be qualitatively observed.}
\label{fig:Scheme}
\end{figure}

The potential profile in Eq. (\ref{eq:TDPotential})
is somewhat idealized. A more realistic
choice should include a Gaussian envelope. For simplicity, we choose to start by considering
a flat-envelope optical lattice, where Bloch's theorem can be invoked with
reasonable confidence. We will see that, remarkably, essentially the same results are obtained when a more realistic Gaussian envelope is used. A sketch of the time-dependent, flat-envelope optical potential and the resulting condensate flow is presented in Fig. \ref{fig:Scheme}.

\section{Preliminary remarks.} \label{sec:Prerem}

We note that the time-dependent amplitude $V(t)$ evolves from $V_{0}\gg\mu_{0}$ at $t \leq 0$ to $V_{\infty}\gtrsim\mu_{0}$ for $t \gg \tau$.
The asymptotic behavior is determined by the initial parameters of the condensate
$(N,g,L)$, the specific form of the final potential ($V_{\infty}$, $d$), and the barrier lowering time scale ($\tau$). The initial potential amplitude, $V_{0}$, plays almost no role provided that it is sufficiently large. More precisely, the condition of initial confinement requires
$\mu_0$ to lie well below the %(extremely narrow) IVAR: not necessary
lowest conduction band of the initial optical lattice potential.
On the other hand, the properties of the final steady state are insensitive to $\tau$ unless $\tau$ is very small (see Subsection \ref{adiabatic}).

The main goal of the present work is to identify the barrier-lowering protocol that best leads to a regime of
quasi-stationary outcoupled flow, by which we mean a flow regime characterized by parameters that vary slowly in time in a sense that will be specified later. As to the space dependence in that regime, we require the density to be as uniform as possible in the region $0<x\alt L$.

In the supersonic region ($x \gtrsim  L+L_{\rm lat}$) we also want a uniform
flow profile, even though, due to the low density,
this will be more difficult to achieve. However, disturbances in the supersonic region
should not affect the spectrum of that part of the Hawking radiation which is emitted into the subsonic region.
On the other hand, in the optical lattice region, the flow should be as close as possible to that of a propagating Bloch wave. At
the boundary between these two regions, large gradients of the
flow speed and density are likely to occur, but the current density should remain essentially uniform.

In what follows, when we refer to bands, we will be meaning the Schrödinger (non-interacting) bands, unless specified otherwise. Bands in a nonlinear context are discussed in Appendices \ref{app:nonlinearol} and \ref{app:olperturbationtheory}.
We will see that, in virtually all cases, the relevant properties of the
optical lattice are determined by its lowest band, provided the lattice is sufficiently long. An important result is that, due to the
finite size of the subsonic reservoir, the quasi-stationary flow is formed only
when the space-averaged chemical potential lands at a value slightly above the bottom
of the lowest lattice band. Because the local chemical potential is approximately uniform along the structure, this implies that the density is small almost everywhere in the optical lattice region. As a consequence, the interaction term can be neglected, since $g|\Psi(x,t)|^{2} \ll (\hbar k_L)^2/m$;
see Appendix \ref{app:nonlinearol} for details.
In the non-interacting regime, Eq. (\ref{eq:TDGP}) becomes the usual Schr\"odinger equation, which for a sinusoidal potential can be transformed into a Mathieu's equation \cite{Abramowitz1988}.

The structure of the final bands can be characterized by the dimensionless parameter
\begin{equation} \label{small-v}
v \equiv mV_{\infty}/8\hbar^2 k_L^{2} \, .
\end{equation}
The nearly-free
atom regime occurs when $v\ll1$.
%as argued in Appendix \ref{app:olperturbationtheory}).
Then bands are wide and
gaps are narrow. By contrast, in the tight-binding regime ($v\gg1$), bands are narrow and widely spaced.
Since $v\propto V_{\infty}d^{2}$,
the band structure can be modified by changing the lattice amplitude
or its spacing.

\section{Ideal optical lattice} \label{sec:numerical}

In this work, the unit length is the initial bulk healing
length $\xi_0$ defined in Section \ref{sec:themodel}.
Accordingly, velocities are measured in units of the sound speed,
$c_{0}\equiv \sqrt{gn_{0}/m}$, and times in units of $t_{0}\equiv \xi_0/c_{0}$. Energies
are expressed in units of the initial chemical potential $\mu_0=mc_{0}^{2}$.
Quasi--one-dimensional condensates of $^{87}$Rb are typically made of $N\sim10^{4}-10^{7}$ atoms and have
a transverse trapping frequency $\omega_{\rm tr}\sim2\pi\times 10^3$ Hz and a confinement length
$L\sim10-400 \mu\text{m}$.
The optical lattice periodicity is bounded
from below, $d>\lambda / 2$nm, for geometrical reasons. The value of lambda is chosen to be sufficiently far from the resonance to avoid any spontaneous emission and on the blue-side of the resonance to produce a repulsive potential ($\lambda < 780$ nm for rubidium atoms).
The simulations are run up to times $t\sim10^4-10^5t_0$. As $t_0=(2n_0a_s\omega_{\rm tr})^{-1}\sim10^{-4}$ s, then $t\sim1-10$ s for our simulations, which is on the order of the mean lifetime of this type of condensates.

In the simulations we use a numerical scheme based on the Crank-Nicolson
method to integrate the time-dependent GP equation (\ref{eq:TDGP}). Hard wall boundary conditions are assumed at $x=0$. At the other end of the finite size computational grid (located at $x=L_{g}$, with $L_{g}$ the total length of the grid), we use absorbing boundary conditions. $L_{g}$ is taken so that the final point of the grid is sufficiently far from the end of the optical lattice for the supersonic region to be clearly observed.
Further details of the numerical method are given in Appendix \ref{app:crnicabc}.

%At the other end of the finite size grid, at distance L_g from the origin, we use absorbing boundary conditions. L_g is taken so that the final point of the grid is sufficiently far from the end of the optical lattice.

\subsection{Analysis of the simulations\label{sub:AnalysisOfSimulations}}

To characterize the quasi-stationary regime, we use the local chemical
potential defined as
%\begin{equation}
%\mu(x,t):=\frac{-\frac{\hbar^{2}}{2m}\frac{\partial^{2}\Psi(x,t)}{\partial x^{2}}}{\Psi(x,t)}+V(x,t)+g|\Psi(x,t)|^{2}\label{eq:LocalChemicalPotential}
%\end{equation}
\begin{equation}
\mu(x,t)\equiv -\frac{\hbar^{2}}{2m}\frac{\partial^{2}\Psi(x,t) / \partial x^{2}}{\Psi(x,t)}+V(x,t)+g|\Psi(x,t)|^{2}\label{eq:LocalChemicalPotential} \, ,
\end{equation}
which can be complex.
For a stationary solution, $\Psi(x,t)=e^{-i\mu t/\hbar}\Psi(x)$,
one has $\mu(x,t)=\mu$, real and independent of $(x,t)$. The current,
%\begin{equation}
%j(x,t)=-\frac{i\hbar}{2m}\left[\Psi^*(x,t)\frac{\partial\Psi(x,t)}{\partial x}-\Psi(x,t)\frac{\partial\Psi^*(x,t)}{\partial x}\right]\label{eq:CurrentFlux}
%\end{equation}
\begin{equation}
j(x,t)=-\frac{i\hbar}{2m}\left(\Psi^*\frac{\partial\Psi}{\partial x}-\Psi\frac{\partial\Psi^*}{\partial x}\right)\label{eq:CurrentFlux}
\end{equation}
is also independent of $(x,t)$ for a stationary solution, as dictated by the continuity equation $\partial_t \rho+\partial_x j=0$. In the quasi-stationary regime the uniformity of $j(x,t)$ is impossible to fulfill strictly, because the current is zero
at $x=0$ while the emitted atoms
carry a nonzero current. Thus, there must be a current gradient and,
by the continuity equation, the density has to be time dependent. The hope is however that, in the
quasi-stationary regime, this time dependence is weak because the condensate leak is slow.

On the other hand, we can expect that, in the quasi-stationary regime, $\mu(x,t)$
is a sufficiently uniform function, with small spatial variations around its space-averaged
mean value.
To check this expectation in a quantitative manner, we introduce the space-averaged
chemical potential $\bar{\mu}(t)$ together with an appropriate measure of its relative spatial fluctuation spread $\sigma(t)$:
\begin{eqnarray}
\bar{\mu}(t) & \equiv  & \frac{\int_{0}^{L_{g}}\mathrm{d}x\,\rho(x,t)\mu(x,t)}{\int_{0}^{L_{g}}\mathrm{d}x\,\rho(x,t)}\nonumber \\
%\sigma(t) & \equiv  & \frac{\sqrt{\frac{\int_{0}^{L_{g}}\mathrm{d}x\,|\Psi(x,t)|^{2}|\mu(x,t)-\bar{\mu}
%(t)|^{2}}{\int_{0}^{L_{g}}\mathrm{d}x\,|\Psi(x,t)|^{2}}}}
%{\bar{\mu}(t)}\label{eq:AverageChemicalPotential}
\sigma(t) & \equiv  & \frac{1}{\bar{\mu}(t)}
\left[
\frac {\int_{0}^{L_{g}}\mathrm{d}x\,\rho(x,t)|\mu(x,t)-\bar{\mu}
(t)|^{2}} {\int_{0}^{L_{g}}\mathrm{d}x\,\rho(x,t)}
\right]^{\frac{1}{2}} \, .
\label{eq:AverageChemicalPotential}
\end{eqnarray}
We recall
that $\mu(x,t)$ and $\bar{\mu}(t)$ can be complex. A nonzero imaginary
part of $\mu(x,t)$ reflects a leaking condensate, as revealed by the local relation
\begin{equation}
%\frac{\partial\rho(x,t)}{\partial t}=\frac{2}{\hbar} \rho(x,t) \text{Im}(\mu(x,t)) \, ,
\frac{\partial\rho}{\partial t}=\frac{2}{\hbar} \rho \, \text{Im} \, \mu \, .
\end{equation}
%\begin{equation}
%\frac{\partial\rho(x,t)}{\partial t}=2\frac{\text{Im}(\mu(x,t))}{\hbar}\rho(x,t)
%\end{equation}
Accordingly, we can define and compute the emission rate per particle  as
\begin{equation}\label{eq:EmissionRate}
\Gamma(t)\equiv \frac{j(L_{g},t)}{\int_{0}^{L_{g}}\mathrm{d}x~\rho(x,t)}=
-\frac{2}{\hbar} \text{Im}\, \bar{\mu}(t) \, ,
\end{equation}
where the continuity equation has been used.
The spatial average of the time-dependent chemical potential [see Eq. (\ref{eq:AverageChemicalPotential})] is mostly determined by the subsonic region, where $\mu(x,t) \simeq g \rho(x,t)$.

A further rescaling of the condensate wave function $\Psi(x,t)\rightarrow\sqrt{n_0}\Psi(x,t)$
reveals more neatly the intrinsic parameters governing the
system. Once the healing length $\xi_0$ is given, only $L/\xi_0,\, d/\xi_0,\,\tau/t_{0},\, V_{\infty}/mc_{0}^{2}$
and $n_{\rm osc}\equiv L_{\rm lat}/d$ (number of oscillations in the optical lattice) are relevant for the problem. We have already noted that
$V_{0}$ plays almost no role in the limit $V_{0}\gg\mu_0$. We find that,
for the pertinent experimental ranges, namely, $L\sim10-400\,\mathrm{\mu m}$, variations of $L/\xi_0$ have little effect on the properties of the quasi-stationary regime.
We have noted that they have a small influence on the time needed to achieve the desired quasi-stationarity, which grows weakly with the initial size of the condensate.
A similar point can be made about $n_{\rm osc}$, which becomes unimportant  when it lies in the range $n_{\rm osc}\sim15-100$. The (relatively small) effect of increasing $n_{\rm osc}$ even further is that there are more spatial fluctuations in the chemical potential, for two reasons: (a) the optical lattice tends to host larger $\mu(x,t)$ spatial fluctuations than the subsonic zone because of atomic reflections across the wells, an effect that is enhanced for larger optical lattices; (b) the larger the lattice, the bigger its contribution to the average chemical potential and its fluctuations.

In summary, except for the above remarks, only the parameters $d/\xi_0,\,\tau/t_{0},V_{\infty}/mc_{0}^{2}$,
have a noticeable effect on the transport properties of the system under study.

\subsubsection{Role of the final band structure} \label{subsub:bandstructure}

\begin{figure}[tb!]
\includegraphics[width=1\columnwidth]{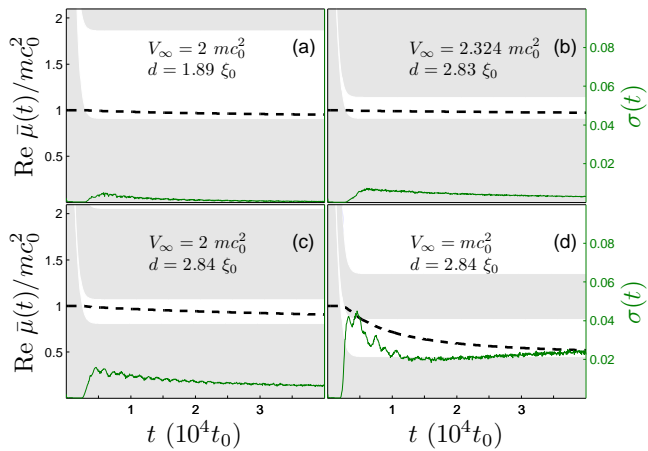} \caption{
Time evolution of the real part of the space-averaged chemical potential $\bar{\mu}(t)$
(black dashed) and its fluctuation spread $\sigma(t)$
(green solid), both defined in Eq. (\ref{eq:AverageChemicalPotential}).
The gap and conduction band of the instantaneous band structure
[computed from the potential Eq. (\ref{eq:TDPotential})] are indicated,
respectively, by grey and white backgrounds.
All graphs are computed with $\tau=500\, t_{0}$,
$L=400\,\mu\text{m}$, $n_{\rm osc}=30$, $N=10^4$, and $\omega_{\rm tr}=2\pi\times4\,\text{kHz}$,
which yields $n_{0}=25\,\mu\text{m}^{-1}$ and $\xi_0=0.3175\, \mu\text{m}$.
The long-time potential amplitude $V_{\infty}$ and the lattice spacing $d$ are indicated
in the graphs. The dimensionless parameter $v$ [see Eq. (\ref{small-v})] takes the values
$(0.0905,0.2350,0.2036,0.1018)$ for graphs (a)-(d).
The setups (a) and (b) are designed to have the same band bottom.
%(a) $d=600$ nm, $V_{\infty}=2\, mc_{0}^{2}$;
%(b) $d=897$ nm, $V_{\infty}=2.324\, mc_{0}^{2}$
%(same value as (a) for the lower limit for the conduction band but
%narrower band). Bottom-left (c): $d=900~\text{nm}$ and $V_{\infty}=2\, mc_{0}^{2}$.
%Bottom-right (d): $V_{\infty}=1\, mc_{0}^{2}$ (same $d$
%as in (c)).
%Second row show the effect of the relative position
%of the $1$ with respect to the lower band.
The simulations are run until a time $4\times 10^4\, t_{0}=5.5\,\mathrm{s}$.
Note that the scale of
$\sigma(t)$ is considerably enlarged.
The initial value $\sigma(0)$ (only observed with some magnification)
is spurious, and is related to the discrete approximation to the derivatives
in Eq. (\ref{eq:LocalChemicalPotential}).}
\label{fig: IdealBandComparison}
\end{figure}

As noted before, the combination of $d$ and $V_{\infty}$
fixes the final band structure. Figure \ref{fig: IdealBandComparison} shows the various
scenarios which one may find depending on the long-time width and position of the lowest band with respect to the
initial chemical potential, $\mu_{0}$. The band structure is computed numerically.
The desired steadiness of the long time behavior improves with the width of the band, as the
first row in Fig. \ref{fig: IdealBandComparison} reveals. In Fig. \ref{fig: IdealBandComparison}a, a favorable
case (wide band) is presented and compared with a less favorable case
in Fig. \ref{fig: IdealBandComparison}b, which has the same conduction band minimum but a narrower band.
After a short transient, a comparison of the relative chemical-potential standard deviation  $\sigma(t)$, as defined in Eq. (\ref{eq:AverageChemicalPotential})
and plotted in this graph, shows a clear advantage in the use of wider
bands. For instance, in Fig. \ref{fig: IdealBandComparison}a, $\sigma(t) \sim 10^{-4}$ in the stationary (long time) regime, about 10 times smaller than in Fig. \ref{fig: IdealBandComparison}b.

Besides, after a transition time of order $\sim 5000 t_0$, all the characteristic magnitudes of the system
shown in Fig. \ref{fig: IdealBandComparison}a vary slowly enough in time to properly view the resulting flow regime
as quasi-stationary. In fact, the leak is so slow that other
processes which limit the lifetime of the condensate (such as condensate decay due to inelastic collisions) operate on a shorter
time scale.

If the chemical potential reaches and goes below the bottom of the conducting band  in a relatively short time, then a transition occurs to an essentially confined
situation where the leaking is exponentially small, corresponding to an atom transmission probability $T(L_{\rm lat})\propto \exp(-\kappa L_{\rm lat})$, where $\kappa \propto \sqrt{E_{\rm min}-\mu}$, with $E_{\rm min}$ the bottom of the conducting band.
This situation whereby one soon reaches the regime $\mu < E_{\rm min}$ is not interesting for our purposes because we need some appreciable
flux in order to form a useful black-hole configuration. In particular, we are typically interested in considering condensates so large that the time needed to reach the bottom of the conduction band is longer than the typical lifetime of the condensate.
%
%When the calculations of Fig. \ref{fig: IdealBandComparison} are repeated for other setups or protocols, such as e.g. one with a smaller $\tau$, we find similar behavior if the chosen interval is $5000 t_0$, but not if we choose $10\tau$ (not shown).

A further argument can be invoked in favor of wide bands. In
virtually all the cases we have addressed, ${\rm Re}\,\bar{\mu}(t)$ drops until it almost reaches the bottom of the conduction band, where leaking
is slow. In the resulting regime, the density in the lattice is very small, $g\bar{n}_r\ll \hbar^2k_L^2/m$, where $\bar{n}_r$ is the mean density in the optical lattice, as defined precisely in subsection \ref{qsr}.
It is shown in Appendix \ref{app:olperturbationtheory} that, in this regime of low interactions, the width of the conduction
band for the linear perturbations (whose evolution is governed by the BdG equations)
is very close to that obtained for the linear Schr\"odinger equation, and the corrections are there computed.
This means that the optical lattice acts like a low-pass
filter, the band width being the equivalent of the cutoff frequency.
The higher the cutoff, the wider is the transmission band of the
lattice. As a consequence,  fluctuations on the subsonic side are transmitted away through the optical lattice, which reduces the space fluctuations in
the chemical potential.

Another trend can be observed in the second row of Fig. \ref{fig: IdealBandComparison}. When placing
$\mu_{0}$ slightly below the top of the conduction band or in the
first gap, as Fig. \ref{fig: IdealBandComparison}d illustrates, the leaking occurs faster but the reached regime presents much larger fluctuations than in the other cases shown in Fig. \ref{fig: IdealBandComparison}. For the purpose of keeping $\sigma(t)\ll 1$, a more favorable situation
for the chemical potential is shown in Fig. \ref{fig: IdealBandComparison}c. There, for the
same length $d$ as in  Fig. \ref{fig: IdealBandComparison}d but a higher barrier amplitude $V_{\infty}$, the chemical potential is initially placed in the final conduction band. This case clearly yields smaller fluctuations,
even though the width of the conduction band is smaller. This shows that, besides having wide bands, one also needs that $\mu_0$ be placed close to the bottom of the final conduction band in order to obtain a more favorable quasi-stationary regime.
This point is further discussed in the next subsection (\ref{qsr}).
%This can be qualitatively explained by the fact that as the difference between $\mu_0$ and the bottom of the conduction band gets bigger, a higher particle flux is needed to reduce the density. This results in larger fluctuations and thus in higher values of $\sigma(t)$.

Within the nearly-free particle approximation ($v\ll 1$), the bottom and top of the first conduction band are given by the relations:
\begin{eqnarray}
E_{\rm min}(v) & =  & 8 E_R (v-v^2+O(v^4)) \nonumber \\
E_{\rm max}(v)& =  & E_R (1+4v-2v^2+O(v^4)) \, ,\label{eq:top-bottom}
\end{eqnarray}
where $E_R \equiv \hbar^2 k_L^2 / 2m$ is the recoil energy of the optical lattice. Given that $k_L=\pi / d$, the condition that the initial chemical potential lies within the final conduction band, i.e.,
\begin{equation}
\label{eq:min-mu-max}
E_{\rm min}(v) < \mu_{0} < E_{\rm max}(v) \, ,
\end{equation}
is guaranteed to be satisfied if
\begin{equation}
\label{eq:m-mu-M}
8E_R v <  \mu_{0} < E_R \, .
\end{equation}
The left inequality is just
\begin{equation}\label{eq:v-min}
\frac{V_{\infty}}{2}<\mu_0~,
%V_{\infty}/2<\mu_0~,
\end{equation}
while the right inequality can be rewritten as:
\begin{equation}
\label{eq:d-max}
d < \frac{\pi}{\sqrt{2}}\xi_0 \, .
\end{equation}
Equations (\ref{eq:v-min}), (\ref{eq:d-max}) express a sufficient condition to satisfy Eq. (\ref{eq:min-mu-max}).

\subsubsection{Non-adiabatic effects} \label{adiabatic}

\begin{figure}[tb!]
\includegraphics[width=1\columnwidth]{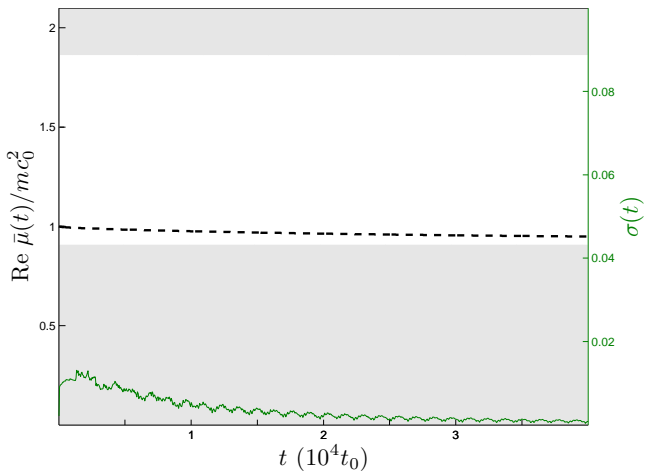} \caption{
Fast barrier lowering.
Same parameters as in Fig. \ref{fig: IdealBandComparison}a except for $\tau=t_{0}$, too short
a time to be observed on this scale.}
\label{fig:IdealLambda1200Tau1Vf2}
\end{figure}

In the favorable situation shown in Fig. \ref{fig: IdealBandComparison}a,
the dependence on $\tau$ is not important as long as
$\tau \gg  t_{0}$.
A simulation is presented in Fig. \ref{fig:IdealLambda1200Tau1Vf2}
which shows that, in the fast regime ($\tau \sim  t_{0}$), and due to the high-frequency excitations
induced by the short lowering time scale,
$\sigma(t)$ remains higher than in the adiabatic case (see Fig. \ref{fig: IdealBandComparison}a).

\subsection{Quasi-stationary regime} \label{qsr}

From the discussion in the previous subsection (\ref{sub:AnalysisOfSimulations}),
and particularly through the situation shown in Fig. \ref{fig: IdealBandComparison}a,
we have learned that, for wide enough bands, initial chemical potential close to the bottom of the final conduction band,
adiabatic evolution ($\tau\gg t_{0}$), and a typical setup, the system evolves towards a quasi-stationary
regime in times shorter than the lifetime of a condensate.

The quasi-stationary regime can be defined as that in which
Re$\,\mu(x,t)$
%[see Eq. (\ref{eq:LocalChemicalPotential})]
is essentially uniform [$\sigma(t)\ll1$] and its global time variations take place
on a time scale of the order of or greater than the condensate lifetime. We focus our study on the most favorable quasi-stationary scenarios, which here we identify with those satisfying $\sigma(t)\lesssim10^{-4}$.

The achievement of this regime is of general interest as a scenario for the study of atom quantum transport. In particular, one may expect spontaneous Hawking radiation to be detectable above a quasi-stationary background with small spatial fluctuations.

In the present section we discuss several features of the condensate
wave function for the quasi-stationary regime. For illustration purposes,
all the graphs considered in this subsection have been obtained
for a system with the parameters of Fig. \ref{fig: IdealBandComparison}a,
which are sufficiently representative.

By writing the condensate wave function as $\Psi(x,t)=\sqrt{\rho(x,t)}e^{i\phi(x,t)}$,
we may introduce two local velocities:
\begin{eqnarray}
v(x,t) & \equiv  & \frac{\hbar\partial_{x}\phi(x,t)}{m} \, ,\nonumber \\
c(x,t) & \equiv  & \sqrt{\frac{g\rho(x,t)}{m}} \, ,\label{eq:SpeedFlowSound}
\end{eqnarray}
$v(x,t)$ being the local condensate flow velocity and
$c(x,t)$ the local speed of sound. The spatial variations of both velocities are small
in the subsonic and supersonic regions, but not in the lattice.
%as argued in
We note that, in that region, $c(x,t)$ must not be regarded as the lattice sound speed; see Appendices \ref{app:nonlinearol} and \ref{app:olperturbationtheory}.

\begin{figure}[tb!]
\includegraphics[width=1\columnwidth]{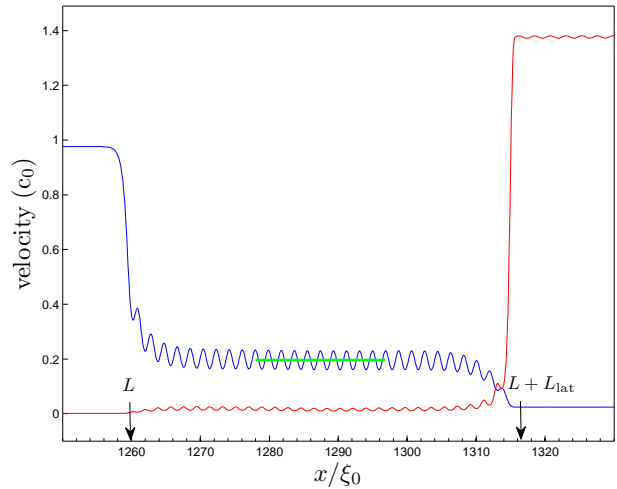} \caption{
Local flow velocity (red) and local speed of sound (blue) at a late time
$t=4\times 10^4\, t_{0}$. The horizontal green segment shows the speed of sound in the optical lattice, computed using (\ref{eq:PerturbativeSound}) with the coefficients there appearing computed numerically. Within this finite lattice the mean density is $\bar{n}_r(t)$, computed by dropping 10 lattice
sites at each end of the lattice.
System parameters are as in Fig. \ref{fig: IdealBandComparison}a.}
\label{fig:IdealCVProfile}
\end{figure}

The profile of both quantities computed at a time, $t=4\times 10^4\, t_{0}$, after a barrier removal time of $\tau=500 t_0$,
is shown in Fig.
\ref{fig:IdealCVProfile}. The subsonic zone shows an essentially flat (uniform) density and flow speed profile in the sense that the spatial fluctuations are on the order of $\sim 10^{-4} n_0$ for the density and $\sim 10^{-3} c_0$ for the flow speed, too small to be observed in Fig. \ref{fig:IdealCVProfile}. In Appendix \ref{app:GPstationary}, an approximate analytical formula [Eq. (\ref{eq:effectivepsi0})] is given for the wave function of the confined condensate which fits the numerical results within this level of accuracy. This good agreement reflects the low value of the flow velocity in the condensate region.

\begin{figure}[tb!]
\includegraphics[width=1\columnwidth]{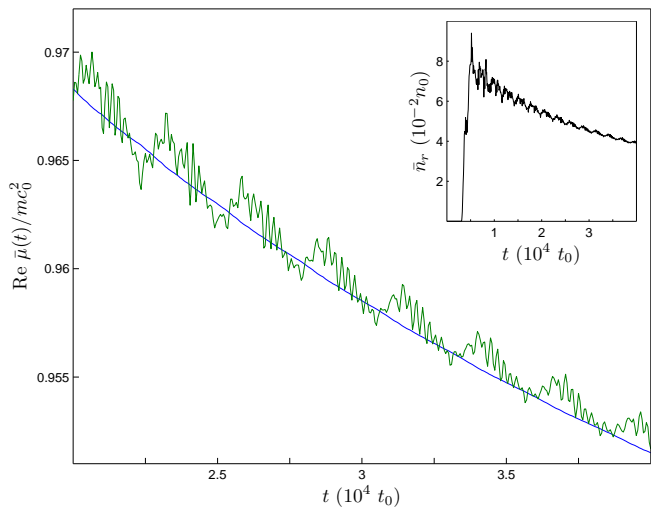}
\caption{System parameters as in Fig.
\ref{fig: IdealBandComparison}a.
Blue: real part of the space-averaged chemical potential $\bar{\mu}(t)$
[see Eq. (\ref{eq:AverageChemicalPotential})]. Green: chemical potential for zero Bloch momentum
computed from the results of Appendix \ref{app:olperturbationtheory} and using
$\bar{n}_{r}(t)$ as the mean density, which is precisely defined in the previous figure and
plotted in the present inset.
}
\label{fig: LatticeFig}
\end{figure}

On the other hand, in the deep central region of the optical lattice,
Bloch's theorem is satisfied. We introduce the space-averaged density $\bar{n}_{r}(t)$ by averaging $\rho(x,t)$ over the optical lattice after excluding 10 lattice sites at each end of the lattice.
That average density, combined with numerically computed quantities that depend on the optical lattice potential, yields an effective sound velocity [see Eq. (\ref{eq:PerturbativeSound})] that is plotted as a horizontal green segment spanning the averaged region in Fig. \ref{fig:IdealCVProfile}.
It can be clearly seen that, within the optical lattice, the flow is subsonic, the horizon lying on its right edge.
In the quasi-stationary regime, $\bar{n}_{r}(t)$
decreases at a rate comparable to the inverse lifetime of the condensate,
as the inset in Fig. \ref{fig: LatticeFig}  shows.

At the edges of the
lattice there are strong variations of the density due to the
the matching between the vastly different densities found on both the
subsonic and the supersonic side.

A check on the approximate validity of Bloch's theorem in the presence of non-linear corrections, for the central part of the optical lattice
and in the quasi-stationary regime, is also shown in Fig. \ref{fig: LatticeFig}.
In this graph, the real part of $\bar{\mu}(t)$
is compared with the
time-dependent chemical potential computed using $\bar{n}_{r}(t)$ and assuming zero Bloch momentum, as explained in the first paragraphs of Appendix \ref{app:nonlinearol}.
%discussion leading to Eq. (\ref{eq:NonLinearOLQuadraticSum}).
The good agreement between the two curves suggests that the condensate is flowing with a very small Bloch momentum.

\begin{figure}[tb!]
\includegraphics[width=1\columnwidth]{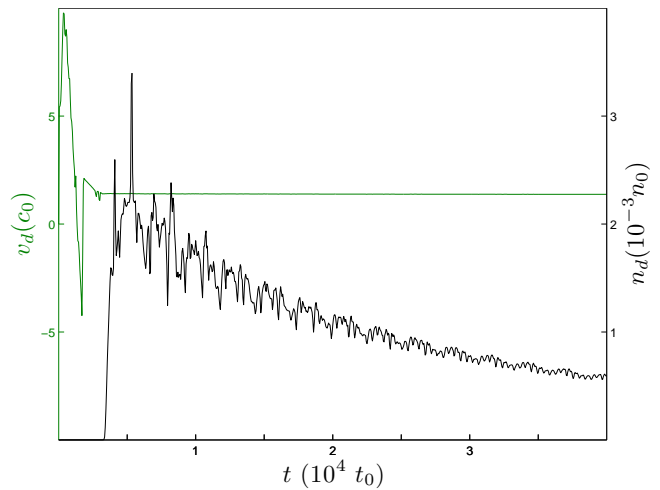} \caption{Time evolution of the mean density and the mean flow
velocity on the supersonic side. The means are taken over the entire downstream region.
System parameters are as in Fig. \ref{fig: IdealBandComparison}a.}
\label{fig:IdealTimenvD}
\end{figure}

In the supersonic zone, once in the quasi-stationary regime,
%and up to the boundary of the numerical grid, $x=L_{g}$,
both density and
flow speed profiles are almost uniform. This is hinted at in Fig. \ref{fig:IdealCVProfile} but not shown explicitly.
Part of the non-flat behavior is due to
small spurious reflections; see Appendix D for details. Conservation of the chemical potential
and the near absence of interaction effects in this
zone make the flow speed almost uniform, because the chemical potential
is almost fully transformed into kinetic energy. On the other hand, the supersonic density
decays with time as the reservoir is depleted, but the process is such that, at each instant, the density profile remains essentially uniform (not shown). An
inhomogeneous density profile would show up only on a space scale much larger than that used in this simulation. A comparison of the decaying supersonic density $n_d(t)$ and the nearly constant supersonic side speed, $v_d(t)$, is shown in Fig. \ref{fig:IdealTimenvD} (subindex $d$ stands for 'downstream' region). The emission rate per particle (not shown), as computed from Eq. (\ref{eq:EmissionRate}),
is practically identical to the product $n_d(t) v_d(t)$, except for a numerical factor corresponding to the instantaneous total number of particles,
which in the quasi-stationary regime is practically constant. This emission rate gives us the typical time scale for the
variation of the number of particles of the system, which is approximately the time scale for the variation of $\mu$. In the quasi-stationary regime considered here, it is $\sim 10^{-6}-10^{-7}~t^{-1}_0$, from which we infer that the typical variation time of the chemical potential is $\sim 10^{6}-10^{7}~t_0$, much longer than the lifetime of a condensate $\sim10^{4}~t_0$.

We conclude that the realization of this quasi-stationary regime needs two fundamental ingredients: the existence of a band structure and the presence of interactions. Without a band structure as that provided by the optical lattice, the condensate would continue leaking through the barrier at a fast rate. On the other hand, the presence of interactions (as reflected in the fact that $\partial \mu / \partial n \neq 0$) allows the condensate to stabilize its flow near the bottom of the conduction band. If the interactions in the subsonic region were negligible, the condensate would empty quickly (if $\mu_0$ lied in the final conducting band) or it would remain confined (if $\mu_0$ lied in the final gap).

Such trends can be seen in the accompanying videos that represent simulations for the same parameters as in  Fig. \ref{fig: IdealBandComparison}a except for $L=10 \, \mu\text{m}$ and $N=250$. The qualitative conclusions are similar. In \href{https://www.youtube.com/watch?v=abKDOO5njIo}{Video 1}, we see the time evolution of the density of a condensate confined by an ideal optical lattice of $30$ barriers. We see that the system achieves the desired quasi-stationary regime. On the other hand, in \href{https://www.youtube.com/watch?v=XS1tcpd3aPk}{Video 2}, we introduce a similar potential but with just a single barrier. We observe that the fluid leaks faster through the single barrier because there is no structure providing a conduction band whose lower boundary is raised close to the chemical potential in order to efficiently slow down the density decrease. This conclusion applies to the class of setups we are considering, which include an initially confined condensate. In other approaches, such as that of Ref. \onlinecite{Kamchatnov2012}, the condensate is projected onto a potential barrier
%(similar to that of \href{https://www.youtube.com/watch?v=XS1tcpd3aPk}{Video 2})
and a quasistationary black-hole regime is also eventually reached.

%We have chosen a scheme based on the deconfinement of a condensate at rest because it seems easier to achieve in the laboratory within the present experimental techniques. In particular, Sec. \ref{sec:Gaussian-shaped} is devoted to the study of a realistic experimental configuration involving a Gaussian-shaped optical lattice. Another potential advantage of the protocol presented in this work is the absence of soliton emission, which could hinder the detection of the quantum quasiparticle signal.

The interaction plays the additional role of providing relaxation channels whereby the condensate lowers its energy while some collective modes are excited. The existence of Landau instabilities (see Appendix \ref{app:nonlinearol}) when $\mu_0$ lies well above $E_{\rm min}$ can be clearly observed in the upper right corner of Fig. 5 of Ref. \onlinecite{Wu2003}, whose chosen parameters are similar to those of the present work. The low value of the critical velocity helps to understand the small value of the condensate Bloch momentum which we infer from the numerical results shown Fig. \ref{fig: LatticeFig} of our present work. The appearance of instabilities can also be viewed as responsible for the fast lowering of the chemical potential after being initially prepared above the final conduction band, as shown in Fig. \ref{fig: IdealBandComparison}d. This interpretation is consistent with the relatively large values found for $\sigma(t)$ when $\mu_0$ is considerably above $E_{\rm min}$.

\section{Gaussian-shaped optical lattice}
\label{sec:Gaussian-shaped}

Here we perform the same analysis as in the previous section
but using a more realistic optical lattice which includes a Gaussian envelope
\cite{Fabre2011,Cheiney2013EPL,Cheiney2013PRA}:
\begin{equation}
%V(x,t)=V(t)e^{-2\frac{(x-L)^{2}}{\tilde{\gamma}^{2}}}\cos^{2}(k(x-L))\label{eq:actualpotential}
V(x,t)=V(t)\cos^{2}\left[k_L(x-L)\right]
\exp\left[-2\left(\frac{x-L}{\tilde{w}}\right)^2\right]
\label{eq:actualpotential}
\end{equation}
where $\tilde{w}=w/\cos(\theta/2)$ (with $w$ the laser beam width and $\theta$ the angle between the laser beams)
plays the role of an effective lattice length.
The time dependence of $V(t)$ is the same as in Eq. (\ref{eq:TDPotential}). Usually, $\tilde{w}$ varies in a range $10-200~\mu\text{m}$. Here, $L$ is the position of the maximum of the lattice Gaussian envelope.
For consistency, we replace the hard wall at $x=0$ by a Gaussian barrier of the type $V_L(x)=U\exp(-2x^2/w_L^2)$ with $w_L=2~\mu\text{m}$ and $U\gg \mu_0$ in order to simulate a more realistic confinement on the left side. This time-independent potential must be added to the time-dependent potential (\ref{eq:actualpotential}) which provides confinement on the right; see Appendix \ref{app:GPstationary} for a detailed description of the initial confinement.

\begin{figure}[tb!]
\includegraphics[width=1\columnwidth]{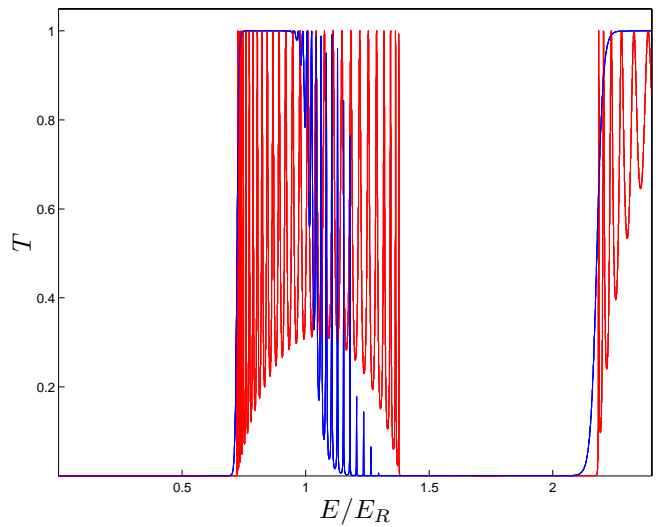}
\caption{
Single atom transmission probability $T(E)$, as a function of energy, for a realistic (Gaussian-shaped) optical lattice
(blue) with the instantaneous value $V(t)=1.6 E_R$ [see Eq. (\ref{eq:actualpotential})] and $\tilde{w}=30 d$, with $E_R$ defined after Eq. (\ref{eq:top-bottom}), and
for an ideal (flat) optical lattice (red) with same amplitude $V(t)$ and $n_{\rm osc}=30$.
}
\label{fig:RealisticIdT}
\end{figure}

\begin{figure}[tb!]
\includegraphics[width=1\columnwidth]{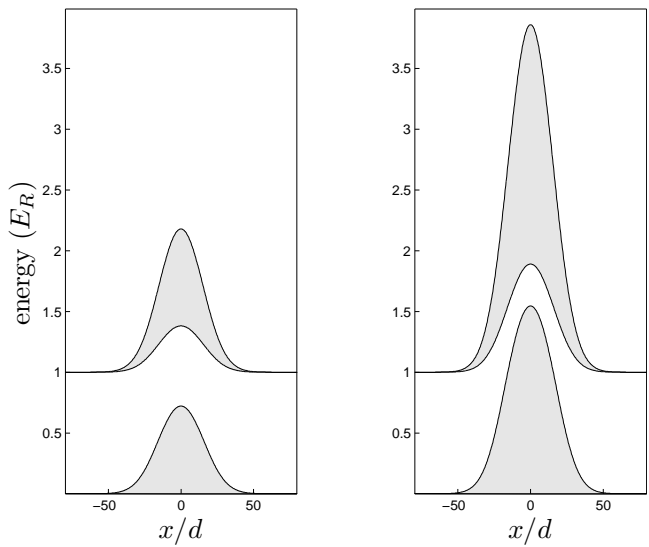}
\caption{Plot of the spatially dependent energy bands for a realistic optical lattice with $\tilde{w}=30 d$. We use the same color criterion as for the band structure of Fig. \ref{fig: IdealBandComparison}. Left panel: the instantaneous value of the amplitude is $V(t)=1.6 E_R$, which corresponds to the case of Fig. \ref{fig:RealisticIdT}. Right panel: the instantaneous value of the amplitude is $V(t)=4 E_R$. In this case, the first conduction band becomes ineffective, as can be expected from the plot, since transmission is always hindered somewhere for the energies of interest.
}
\label{fig:SpaceBands}
\end{figure}

In the ``adiabatic'' regime ($\tilde{w}\gg d$), the solutions of the linear Schr\"odinger equation
for this type of potentials can show features similar to those found for an ideal optical
lattice with the same instantaneous amplitude $V(t)$, as can be seen in Fig. \ref{fig:RealisticIdT}, where the transmission bands are compared.
If we focus on the long-time limit ($V(t)=V_{\infty}$), the realistic potential acquires the form
\begin{equation}
V(x)=V_{\infty}(x)\cos^{2}\left[k_L(x-L)\right] \, ,
\end{equation}
where
\begin{equation}
V_{\infty}(x)=V_{\infty}\exp\left[-2\left(\frac{x-L}{\tilde{w}}\right)^2\right]
\label{V-infty-x}
\end{equation}
is a slowly varying function. Then, we have a locally ideal optical lattice at each point of the space with amplitude $V_{\infty}(x)$. Bloch's theorem can also be applied locally and a local band structure results which is plotted as a function of space in Fig. \ref{fig:SpaceBands}. A similar type of reasoning was already used in Refs. \cite{Carusotto2000,Santos1998a,Santos1999}. The left panel presents the setup whose single atom transmission is plotted in Fig. \ref{fig:RealisticIdT}. Since the bottom of the lowest lattice conduction band is an increasing function of the periodic potential amplitude, the bottleneck for transmission across the realistic lattice occurs at the center of its Gaussian envelope. This fact explains the accurate coincidence between the bottom of both conduction bands shown in Fig. \ref{fig:RealisticIdT}. We also see that, for $E>E_R$ [defined after Eq. (\ref{eq:top-bottom})], the particle encounters a gap somewhere along the Gaussian lattice, and this explains why in Fig. \ref{fig:RealisticIdT} the transmission begins to decay for $E>E_R$. For $E_{\rm min}(v)<E<E_R$, the setup shows a plateau of essentially perfect atom transmission. The absence of interference oscillations in this region is due to the adiabatic variation of the lattice envelope. The right panel presents the different case of $E_R<E_{\rm min}(v)$. From the foregoing arguments, we expect not to find a conduction band, as can be numerically confirmed. We conclude that, in order to have a well defined conduction band for the realistic lattice, the condition $E_R>E_{\rm min}(v)$ is required, which implies $V_{\infty}<2.33~E_R$. Combining all these considerations, the conclusion is reached that a necessary condition for achieving a quasi-stationary regime is $E_{\rm min}(v)<\mu_0$. We also require $\mu_0 < E_R$ to avoid having $\mu_0$ lying too high above $E_{\rm min}(v)$, which, as found for the ideal lattice, tends to generate relatively high values of $\sigma(t)$. This last inequality is equivalent to Eq. (\ref{eq:d-max}).

Here space can also be divided into three zones. In the quasi-stationary regime, both the subsonic and supersonic zones are located where
%the real part of the chemical potential, $\text{Re}~(\mu(x,t))$
the Gaussian envelope amplitude is negligible compared to the chemical potential, i.e., where
%$|V(t)| \ll \text{Re} \mu(x,t)$
%$V_{\infty}\exp[-2(x-L)^2/\tilde{w}^2]\ll {\rm Re} \, \bar{\mu}(t)$
\begin{equation}
V_{\infty}(x) \ll \, {\rm Re} \, \bar{\mu}(t)
\label{eq: v-inf-ll-mu}
\end{equation}
[see Eqs. (\ref{eq:AverageChemicalPotential}) and (\ref{V-infty-x})]. In order for the subsonic side to be well differentiated, we set $L\gg \tilde{w}$. The optical lattice region is the complementary of the subsonic and supersonic zones, i.e., the region where (\ref{eq: v-inf-ll-mu}) does not apply.

\begin{figure}[tb!]
\includegraphics[width=1\columnwidth]{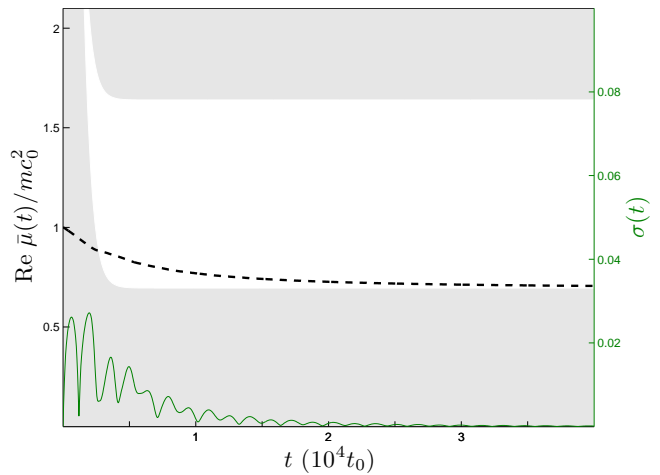} \caption{Time evolution of the real part of the chemical potential and its fluctuation spread in a realistic (Gaussian-shaped) optical lattice.
The parameters are $\tilde{w}=50~\mu\text{m}$, $d=600~\text{nm}$, $\tau=500~t_{0}$, $V_{\infty}=1.5~ mc_{0}^{2}$ and we have taken $\xi_0=0.3053~\mu\text{m}$. The confinement parameters are $N=9161$, $L=420~\mu\text{m}$, and $\omega_{\rm tr}=2\pi\times 4~\text{kHz}$.}
\label{fig:RealisticBand}
\end{figure}

The requirements of quasi-stationarity are similar to those formulated for the ideal optical lattice. Specifically, the quasi-stationary regime
requires broad conduction bands, an initial chemical potential close to the bottom of the final conduction band,
and a barrier amplitude that evolves not very fast. We also find that the condensate leaks relatively fast until ${\rm Re} \bar{\mu}(t)$ [Eq. (\ref{eq:AverageChemicalPotential})] approaches the bottom of the conduction band. All these features can be observed in Fig. \ref{fig:RealisticBand}, which is the Gaussian-envelope equivalent of Figs. \ref{fig: IdealBandComparison}-\ref{fig:IdealLambda1200Tau1Vf2}. We reach a quasi-stationary state in which $\sigma(t)\sim 10^{-4}$.  The bands in Fig.  \ref{fig:RealisticBand} are computed as in the ideal case, assuming a uniform barrier amplitude $V(t)$. As noted when discussing Fig. \ref{fig:RealisticIdT}, the positions of the bottom of the ideal and the realistic conduction (or transmission) bands are very similar, so the lower threshold of the transmission band can still be a good reference value to discuss the evolution of $\bar{\mu}(t)$.

We notice that in the Gaussian case, the condensate apparently leaks from the beginning of the simulation. What is actually happening is that the chemical potential is already lowered by the initial expansion of the condensate towards the neighboring, low-amplitude region of the Gaussian optical lattice, even when the leaking (towards the right side of the Gaussian envelope) is not yet occurring. This process can be observed in the simulation later presented in \href{https://www.youtube.com/watch?v=b9YA-Efd4F8}{Video 3} at the end of Section \ref{sub:location-sonic-horizon}. The situation contrasts with that shown in Fig. \ref{fig: IdealBandComparison}, where the condensate only begins to leak when the chemical potential is placed within the conduction band.

\begin{figure}[tb!]
\includegraphics[width=1\columnwidth]{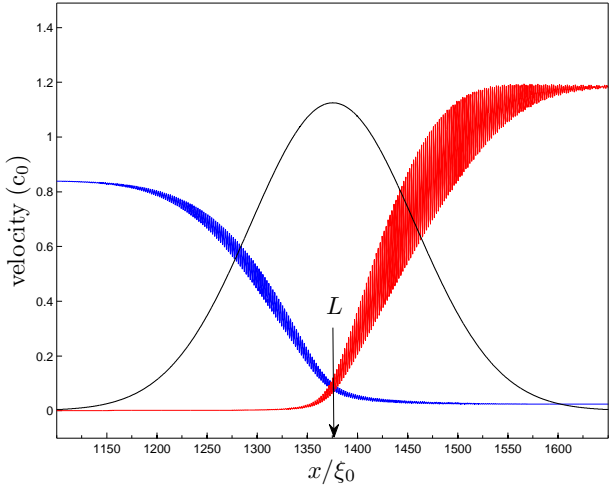} \caption{
Local flow velocity (red) and local speed of sound (blue)
at $t=4\times 10^4\, t_{0}$. System parameters are as in Fig. \ref{fig:RealisticBand}.}
\label{fig:RealisticCVProfile}
\end{figure}

\begin{figure}[tb!]
\includegraphics[width=1\columnwidth]{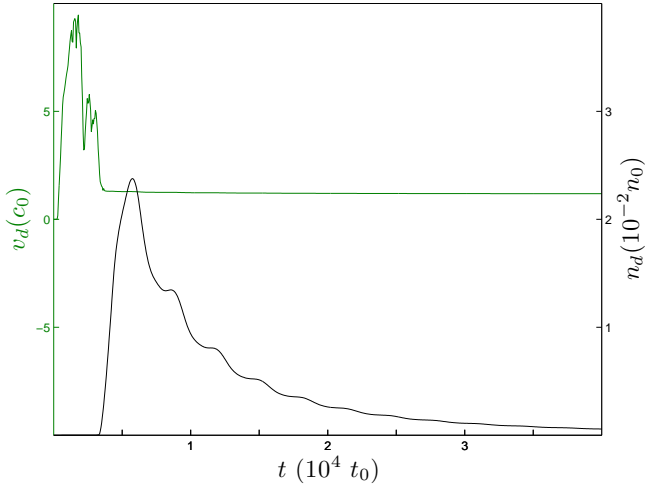} \caption{
Time evolution of the mean density and the mean flow
velocity on the supersonic side for a condensate emitting through a Gaussian-shaped optical lattice. System parameters are as in Fig. \ref{fig:RealisticBand}}
\label{fig:RealisticTimenvD}
\end{figure}

We also study the corresponding quasi-stationary state. For that purpose, we take a snapshot of the configuration at $t=4 \times 10^4~t_0$ for the parameters in Fig. \ref{fig:RealisticBand}. We compare the profiles of $c(x,t)$ and $v(x,t)$ in Fig. \ref{fig:RealisticCVProfile}, which is the realistic equivalent of Fig. \ref{fig:IdealCVProfile}. The apparently sharper oscillations, as compared to those in Fig. \ref{fig:IdealCVProfile}, are due to the different horizontal scales used. The larger oscillations of the flow velocity beyond the horizon with respect to those inside the lattice subsonic region in Fig. \ref{fig:IdealCVProfile} can be explained because of the large difference in space-averaged flow velocities. In the supersonic region, we find again essentially flat profiles for the density and flow velocity, with their time evolution shown in Fig. \ref{fig:RealisticTimenvD}. The general features of this quasi-stationary configuration are similar to those of the ideal case, but some interesting new features appear.

\subsection{Location of the sonic horizon and related properties \label{sub:location-sonic-horizon}}

We notice in Fig. \ref{fig:RealisticCVProfile} that the horizon seems to be placed at the maximum of the Gaussian envelope.
Actually, this can be explained on quite general grounds by invoking the properties of the quasi-stationary regime and the adiabaticity condition $\tilde{w}\gg d$, which allows us to think in terms of a local band structure stemming from a periodic potential of local amplitude $V_{\infty}(x)$. Then we can use an adiabatically space-dependent version of Eqs. (\ref{eq:oldmu})-(\ref{eq:olhdmu}) (where the sound speed, atom current and chemical potential are obtained for an infinite optical lattice) by making every parameter slowly dependent on $x$.  In particular, we take:
\begin{eqnarray}\label{eq:SEQSS}
%s(x)&=&\sqrt{\frac{gn_r(x)}{m^*(x)}\alpha_{0}^{(1)}(x)}\nonumber \\
s(x)&=& \left[\frac{gn_r(x)}{m^*(x)}\alpha_{0}^{(1)}(x)\right]^{\frac{1}{2}}\nonumber \\
j(x)&\simeq&n_r(x)\bar{v}(x) \\
\mu(x)&\simeq&E_{\rm{min}}(x)+\frac{1}{2}m^*(x)\bar{v}^2(x)+m^*(x)s^2(x)\, , \nonumber
\end{eqnarray}
where we neglect the time dependence because the system is assumed to be already in the quasi-stationary regime.
The local averages for $n_r,\bar{v}$ are taken over several lattice periods. In the quasi-stationary regime, the chemical potential is already close to the bottom of the conduction band, so the perturbative study used in Appendix \ref{app:olperturbationtheory} is valid. Taking spatial derivatives, while noting that the chemical potential is almost uniform, $\mu(x)\simeq \bar{\mu}$, and that $\partial_x j(x)$ can be neglected (as implied by the continuity equation and quasi-stationarity), we arrive at:
\begin{equation} \label{eq:derivative-mu-zero}
0=E'_{\rm{min}}+\frac{1}{2}m^{*'}\bar{v}^2+\frac{\alpha^{(1)'}_0}{\alpha^{(1)}_0}m^*s^2+m^*(\bar{v}^2-s^2)\frac{\bar{v}'}{\bar{v}}\, .
\end{equation}
The quantities $E_{\rm{min}}(x),m^*(x),\alpha_{0}^{(1)}(x)$ depend on $x$ through the amplitude of the envelope, $V_{\infty}(x)$, and they increase with its value, provided that the envelope amplitude is always positive
%[see Eq. (\ref{m-alpha-1}) for $m^*$ and $\alpha_{0}^{(1)}$].
[see Eq. (\ref{m-alpha-1})].
Therefore, the first three terms in the r.h.s. of (\ref{eq:derivative-mu-zero}) have the same sign.

Let us assume that we have a horizon ($s=\bar{v}$) somewhere in the optical lattice. We prove next that a necessary implication is that an envelope maximum or minimum exists at that point. It has just been noted that the first three terms in (\ref{eq:derivative-mu-zero}) have the same sign. Thus, their sum can only be zero whenever the derivative of the amplitude is zero, i.e., when $V'_{\infty}(x)=0$. In our setup, this means that we have an amplitude maximum at the horizon. A proof of a similar result in the case of a single potential barrier, based in a hydrodynamical approximation, was already given in Ref. \onlinecite{Giovanazzi2004}.

Now we consider the inverse implication. Assume we have $V'_{\infty}(x)=0$ (which in our setup is the case at $x=L$). This implies that the first three terms in (\ref{eq:derivative-mu-zero}) are zero. As a consequence, we are left with two possibilities:
\begin{equation} \label{eq:horizon}
s(L)=\bar{v}(L) \,
\end{equation}
(i.e. a horizon) or $\bar{v}'=0$. By the continuity equation, the second option implies a density minimum, which must be ruled out in our current single Gaussian barrier setup. However, it can be a perfectly feasible result in other experimental contexts.

Finally, we note that Eq. (\ref{eq:derivative-mu-zero}) can also be written as
\begin{equation}
0=E'_{\rm{min}}+\frac{1}{2}m^{*'}(\bar{v}^2+2s^2)+m^*(\bar{v}\bar{v}'+2ss')\, ,
\end{equation}
and, as a corollary of the foregoing analysis, we find that, at the horizon, $s'(L)=-\bar{v}'(L)/2$.

We can further exploit the previous results. For example, we can obtain the value of the density and the current at $x=L$ as a function of $\bar{\mu}$:
\begin{eqnarray}\label{eq:njhorizon}
g n(L)&=&\frac{2}{3}\frac{\bar{\mu}-E_{\rm{min}}}{\alpha^{(1)}_0} \, ,\nonumber \\
j(L)&=&n_r(L)\bar{v}(L)=\left(\frac{2}{3}\right)^{\frac{3}{2}}
\frac{\left(\bar{\mu}-E_{\rm{min}}\right)^{\frac{3}{2}}}{g\alpha^{(1)}_0\sqrt{m^*}} \, ,
\end{eqnarray}
which are very good approximations to the actual numerical values. Here the dependence on $L$ of the various parameters is understood. Using (\ref{eq:njhorizon})
we can arrive at a differential equation for the time evolution of $\bar{\mu}$. First we note that, from the continuity equation, we can write:
\begin{equation}\label{eq:Oden}
\frac{dN_L}{dt}=-j(L) \, ,
\end{equation}
where $N_L$ is the number of particles contained between $x=0$ and $x=L$. As the subsonic region is in the Thomas-Fermi regime (see Appendix A), we can take $\bar{\mu}\simeq gN_{\rm{sb}}/L_{\rm{sb}}$, where $N_{\rm{sb}}$ is the number of particles in the subsonic region and $L_{\rm sb}$ its size. As the density in the optical lattice is small, we can assume $N_L\simeq N_{\rm{sb}}$ (which implies $\bar{\mu} \propto N_L$) and write Eq. (\ref{eq:Oden}) as:
\begin{equation}\label{eq:odemu}
\frac{d\bar{\mu}}{dt}=-C\left(\bar{\mu}-E_{\rm{min}}\right)^{\frac{3}{2}} \, ,
\end{equation}
where $C$ is a positive constant independent of $\bar{\mu}$. The solution of this equation is
\begin{equation}\label{eq:mute}
\bar{\mu}(t)=E_{\rm{min}}+\frac{4}{C^2(t-t_1)^2} \, ,
\end{equation}
(with $t_1$ an integration constant), which fits the numerical data of Fig. (\ref{fig:RealisticBand}) reasonably well.

Finally, we can estimate the value of the Hawking temperature, which is given by:
\begin{equation}\label{eq:THexp}
k_BT_H=\frac{\hbar}{2\pi}\frac{d}{dx}\left[\bar{v}(x)-s(x)\right]_{x=L}
\end{equation}

If we note that we operate in the nearly-free atom approximation ($v \ll 1$) and in the weak interaction regime ($gn_r \ll E_R$), and derive twice the third equation (\ref{eq:SEQSS}), we obtain
\begin{equation}\label{eq:TH}
k_BT_H \simeq \frac{\hbar}{2\pi\tilde{w}}\sqrt{\frac{3V_{\infty}}{m^*}}(1-v) \, ,
\end{equation}
which gives a good estimate of the numerical value of the Hawking temperature. Noting that $V_{\infty}\sim\mu_0,m^*\sim m$, we obtain $k_BT_H\sim\xi_0\mu_0/\tilde{w}\sim10^{-2}\mu_0\ll\mu_0$. The temperature of the condensate is typically of the order of $\mu_0/k_B$, so we conclude $T_H\sim10^{-2}T\ll T$. Similar estimations for the value of the Hawking temperature were already given in Ref. \cite{Wuster2007}. 

To observe the birth of the black hole and to check that the horizon position naturally evolves towards the maximum of the optical lattice envelope, we have created a movie (\href{https://www.youtube.com/watch?v=b9YA-Efd4F8}{Video 3}) that shows the time evolution of the coarse-grained velocities $(\bar{c},\bar{v})$ of the emitting condensate using the setup parameters of Fig. \ref{fig:RealisticBand}. At long times the predicted coincidence between the sonic horizon and the maximum of the Gaussian envelope in the stationary regime can be clearly observed.

When applied to an ideal optical lattice, the above arguments on the position of the horizon yield no preferred point for the location of the horizon because the envelope is uniform. Actually, in the bulk of the lattice, since $V'_{\infty}=0$ everywhere, the natural outcome [from the discussion leading to Eq. (\ref{eq:horizon})] is $\bar{v}'=0$ everywhere, i.e., the mean velocity and, by quasi-stationarity, the mean density are also uniform, as can be observed in Fig. \ref{fig:IdealCVProfile}. This fact only leaves two options: either the lattice bulk is subsonic or it is supersonic. The latter choice is energetically unstable (see Ref. \onlinecite{Wu2003}) and, as a consequence, the subsonic regime is energetically favored in the bulk of the lattice. In the rightmost region, where the potential is not present, the flow has to be supersonic, so the only
possibility for the horizon is to lie at the right extreme of the lattice, as can be seen in Fig. \ref{fig:IdealCVProfile}.
%We stress the fact that all the added results obtained in this subsection (Eqs. \ref{eq:njhorizon}-\ref{eq:TH}) rely on the special position of the horizon in the Gaussian-shaped case, and that none of them applies to the ideal lattice profile.

\section{Conclusions} \label{sec:conclusions}

Within a mean-field description, we have investigated the process whereby an initially confined atom condensate is coherently outcoupled as the barrier on one side is gradually lowered. The goal has been to identify the barrier-lowering protocol which best leads to a quasi-stationary sonic black hole located at the interface between subsonic and supersonic flow. We find that the use of an optical lattice for the lowered barrier is convenient to achieve a regime of quasi-stationary flow with minimal value of the fluctuation spread. First we have focused on an optical lattice of finite length and uniform amplitude. We find that the long-time band structure of the optical lattice greatly influences the asymptotic behavior of the emitted atom flow. Within this class of setups, the best quasi-stationary flow is achieved when the lowest conduction band is broad and the initial chemical potential lies not much high above the bottom of the final conduction band. In the optimal cases, the relative value of the spatial fluctuations can be as small as $\sigma(t)\sim 10^{-4}$. When we replace the uniform amplitude of the optical lattice by a more realistic Gaussian envelope, we find that the results are similar to those of a uniform lattice with the amplitude of the envelope maximum. Quite interestingly, we argue analytically and check numerically that, in the quasi-stationary regime, the horizon separating the regions of subsonic and supersonic flow is pinned down right at the Gaussian maximum. We also find that the Gaussian envelope is quite efficient in guaranteeing a small deviation from the ideal stationary flow.

Whether the quasi-stationary regimes here identified can become scenarios for the detection of Hawking radiation, is something that will have to be confirmed by a future study of the quasiparticle dynamics operating against the background of seemingly favorable mean-field configurations.

\acknowledgments
We thank I. Carusotto and R. Parentani for valuable discussions. This work has been supported by MINECO (Spain) through grants FIS2010-21372 and FIS2013-41716-P,
Comunidad de Madrid through grant MICROSERES-CM (S2009/TIC-1476), and the Institut Universitaire de France.

\appendix
%dummy comment inserted by tex2lyx to ensure that this paragraph is not empty

%\section{Stationary solution of GP equation.}
\section{Initial configuration of the condensate.}

\label{app:GPstationary}

In this appendix, we compute the initial profile of the condensate, which
at early times ($t<0$) experiences a confining time-independent potential.
We require a hard-wall boundary condition at $x=0$, which implies, via continuity equation, that the phase of the condensate is constant in space. The amplitude $A(x)\equiv |\Psi(x)|$ of the
solution to the time-independent Gross-Pitaevskii (GP) equation Eq. (\ref{eq:GP})
in the region where there is no potential, satisfies the equation
\begin{equation}
\left(-\frac{\hbar^{2}}{2m}\partial_{x}^{2}+gA^{2}\right)A=\mu_{0}A \, .
\end{equation}
As is well known, this equation can be interpreted as the equation of motion for a particle with ``coordinate''
$A$ and ``time'' $x$ in a certain potential
\begin{eqnarray} \label{eq:GP-potential}
W(A) & = &  \tilde{\mu}_{0}A^{2}-\frac{\tilde{g}}{2}A^{4} \nonumber \\
\tilde{\mu}_{0} & = & \frac{m}{\hbar^{2}}\mu_{0},~\tilde{g}=\frac{m}{\hbar^{2}}g \, .
\end{eqnarray}
Invoking
``energy'' conservation, the equation can be integrated as
%\begin{eqnarray}
%\frac{1}{2}A'^{2}+W(A) & = & E_{A}\nonumber \\
%\tilde{\mu}_{0} & = & \frac{m}{\hbar^{2}}\mu_{0},~\tilde{g}=\frac{m}{\hbar^{2}}g \,
%\end{eqnarray}
\begin{equation}
\frac{1}{2}A'^{2}+W(A) = E_{A} \, ,\label{eq:effectivenergyconservation}
\end{equation}
where $E_A$ is the total energy of this effective motion.

Following Ref. \onlinecite{Zapata2011}, Eq. (\ref{eq:effectivenergyconservation}) can be rewritten
in terms of $\rho(x)=A^2(x)$ as
\begin{equation}
\rho'^{2}=4\tilde{g}(\rho-e_{1})(e_{2}-\rho)(e_{3}-\rho),
\end{equation}
where
\begin{equation}
0=e_{1}\leq\rho\leq e_{2}<e_{3},
\end{equation}
and $e_{2,3}$ are the zeros of
\begin{equation}
\rho^{2}-2\frac{\tilde{\mu}_{0}}{\tilde{g}}\rho+2\frac{E_{A}}{\tilde{g}}=0.
\label{eq:ellipticalpolynomial}
\end{equation}

The solution can be expressed in terms of elliptic functions \cite{Abramowitz1988}. Imposing
the boundary condition $\rho(0)=0$, one obtains a solution of the form
\begin{equation}\label{eq:generalGPsolutionin}
\rho(x)=e_{2}\text{sn}^{2}(\sqrt{\tilde{g}e_{3}}x,\nu),~\nu=\frac{e_{2}}{e_{3}}\, .
\end{equation}
In order to determine $e_{2,3}$, another boundary condition is needed, together with the particle number normalization $\int\mathrm{d}x\, \rho(x)=N$. From (\ref{eq:GP-potential}) and (\ref{eq:ellipticalpolynomial}), the chemical potential can be written as
\begin{equation}
\mu_{0}=g\frac{e_{2}+e_{3}}{2}~.
\label{eq:muzeros}
\end{equation}

\subsection{Ideal confinement}

The ideal confinement boundary condition is defined as $A(L)=0$, and the condensate is confined between $0$ and $L$.
Using Eq. (\ref{eq:generalGPsolutionin}) we find
\begin{equation}
\sqrt{\tilde{g}e_{3}}L=2nK(\nu), \, \, n\in\mathbb{N} \, ,
\label{eq:nquantization}
\end{equation}
where $K(\nu)$ is the complete elliptic integral of the first kind \cite{Byrd1971}.
%\cite{Abramowitz1988}.

Hereafter we work with the ground state ($n=1$). The particle number normalization is
\begin{equation} \label{eq:N-integr}
N=\int_{0}^{L}\mathrm{d}x~e_{2}~\text{sn}^{2}\left(\sqrt{\tilde{g}e_{3}}x,\nu\right).
\end{equation}
By performing the integral in Eq. (\ref{eq:N-integr}) and using (\ref{eq:nquantization}) we have
\begin{equation}
N=\frac{e_{2}}{\sqrt{\tilde{g}e_{3}}}\frac{2}{\nu}\left[K(\nu)-E(\nu)\right],\label{eq:Nellipticalintegral-1}
\end{equation}
where $E(\nu)$ is the complete elliptic integral of the second kind \cite{Byrd1971}. Equations (\ref{eq:nquantization}) and (\ref{eq:Nellipticalintegral-1}) lead to
\begin{equation}
N\tilde{g}L=4K(\nu)\left[K(\nu)-E(\nu)\right],\label{eq:Nellipticalintegral-2}
\end{equation}
or
\begin{equation}
4K(\nu)\left[K(\nu)-E(\nu)\right]=\left(\frac{L}{\xi}\right)^{2}\, ,\label{eq:mequation}
\end{equation}
where the healing length
%$\xi\equiv \sqrt{\frac{\hbar^{2}L}{mgN}}$
$\xi\equiv \sqrt{\hbar^{2}L/mgN}$
is not identical to $\xi_0$ defined in section \ref{sec:themodel}.

After eventually solving for $\nu$, $e_{2}$ and $e_{3}$,  Eq. (\ref{eq:generalGPsolutionin}) can be rewritten as
\begin{equation}
A(x)=\sqrt{e_{2}}~\text{sn}\left(2K(\nu)\frac{x}{L},\nu\right)\label{eq:idealconfinementsolution}
\, .
\end{equation}

For the chemical potential, we obtain, using (\ref{eq:muzeros})-(\ref{eq:nquantization})
\begin{equation}
\mu_{0}=\frac{2\hbar^2}{mL^2}(1+\nu)\left[K(\nu)\right]^2~.
\label{eq:muelliptic}
\end{equation}

\begin{figure}[tb!]
\includegraphics[width=1\columnwidth]{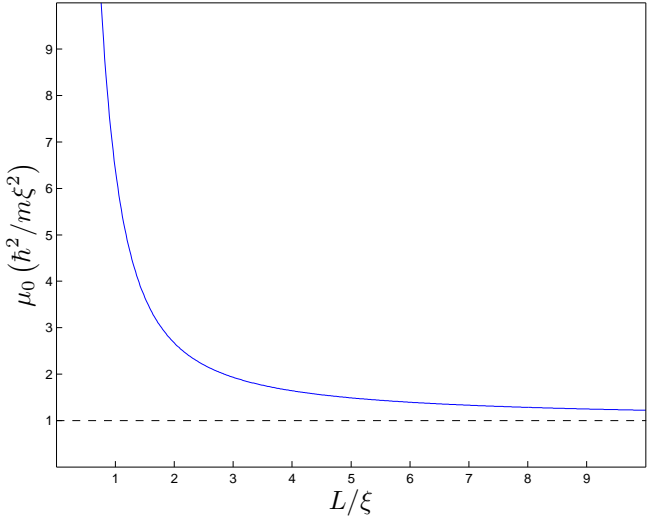} \caption{Computation of the chemical potential as a function of $L/\xi$ using Eqs. (\ref{eq:mequation}) and (\ref{eq:muelliptic}), for an ideal lattice confined between hard walls and in equilibrium. When $L/\xi\ll1$, we are in the Schrödinger limit in which $\mu_0\sim 1/L^2$ and when $L/\xi\gg1$, we are in the Thomas-Fermi regime and then $\mu_0\simeq\hbar^2/m\xi^2$.}
\label{fig:MuLxi}
\end{figure}

Taking into account that $\nu$ is a function of $L/\xi$, as given by Eq. (\ref{eq:mequation}), we plot Eq. (\ref{eq:muelliptic}) in Fig. \ref{fig:MuLxi}. In order to find $\nu$, Eq. (\ref{eq:mequation}) must be solved numerically. However, good approximate solutions  can be found. We can clearly distinguish two different regimes:  $L\ll\xi$ and  $L\gg\xi$. The physical interpretation is straightforward because
the ratio between the kinetic energy and the interaction energy is
\begin{equation}
\frac{E_{\rm int}}{E_{\rm kin}}
%\sim\frac{gN/L}{\hbar^2/mL^2}\sim\frac{\frac{\hbar^2}{m\xi^2}}{\frac{\hbar^2}{mL^2}}=\left(\frac{L}{\xi}\right)^2
\sim\frac{gN/L}{\hbar^2/mL^2}\sim\frac{\hbar^2/m\xi^2}{\hbar^2/mL^2}=\left(\frac{L}{\xi}\right)^2 \, .
\end{equation}

Then, $L\ll\xi$ is the Schrödinger limit in which we have $\nu\simeq0$. In that limit we arrive at the well-known result $\sqrt{\tilde{g}e_{3}}L=\pi$ and $\rho(x)=e_2~\sin^2(\pi x/L)$. In all the cases considered in this work, $L\gg\xi$, so we work in the limit in which interactions represent the main contribution to the chemical potential (Thomas-Fermi regime). $K(\nu)$ diverges when $\nu\rightarrow1$
while $E(\nu)$ remains finite. From (\ref{eq:mequation}) this means that $\nu\simeq1$. In
fact, there are cases in which $1-\nu$ is so small that it falls below computer floating-point relative accuracy. In those cases, the only way to obtain the solution is through asymptotic expansion. One can prove that, in that limit \cite{Byrd1971},
\begin{equation}
K(\nu)\simeq\ln\frac{4}{\sqrt{1-\nu}},~E(\nu)\simeq 1.
\end{equation}
Thus, Eq. (\ref{eq:mequation}) is rewritten as
\begin{equation}
K^{2}-K-\frac{r^{2}}{4}=0,~r=\frac{L}{\xi},
\end{equation}
and from its solution we get $2K=1+\sqrt{1+r^{2}}$. Therefore, $1-\nu=16e^{-2K}\ll1$ which implies both
$e_{2}\simeq e_{3}$ and $\mu_{0}\simeq ge_{2}$. Equation (\ref{eq:nquantization})  implies
\begin{equation}
e_{3}=\frac{4K^{2}}{r^{2}}\frac{N}{L}\simeq\left(1+2\frac{\xi}{L}\right)\frac{N}{L}\, .
\end{equation}
and then
\begin{eqnarray}\label{eq:muxi}
\frac{\mu_{0}}{\hbar^2/m \xi^2}
&=&  \frac{n_{0}}{N/L}=\left(\frac{\xi}{\xi_0}\right)^2  =\nonumber \\
&=& 2(1+\nu)\left[\frac{K(\nu)}{r}\right]^2 \nonumber \\
&\simeq& 1+2\frac{\xi}{L}+2\left(\frac{\xi}{L}\right)^2~.
\end{eqnarray}

Collecting all these results, the wave function can be effectively approximated by
\begin{equation}
\Psi_{0}(x)\equiv \left\{ \begin{array}{cc}
\sqrt{e_{2}}\tanh\left(2K\frac{x}{L}\right), & 0\leq x\leq\frac{L}{2}\\ \\
\sqrt{e_{2}}\tanh\left[2K\left(1-\frac{x}{L}\right)\right], & \frac{L}{2}\leq x\leq L
\end{array}\right.\label{eq:effectivepsi0}
\end{equation}
because $\text{sn}(x,1)=\tanh(x)$. The function $\tanh$ quickly
reaches the asymptotic value $1$, which means that in the central zone
of the confinement region, the solution is essentially flat.

\subsection{Ideal optical lattice potential}

For computational purposes, the initial rightmost boundary condition is also taken $A(L)=0$ but now half a period of the lattice potential lies inside the confinement region, as explained in the main text. This artificial boundary condition does not create a problem because we take $V_{0}\gg gN/L$ so the function inside the lattice potential is exponentially small. In order to compute the stationary solution in the region where the potential is present, a numerical solution of the GP equation has to be performed. In the situations considered in the present work, $L\gg d$, so the wave function is very similar to that of the ideal confinement case. For numerical convenience, instead of fixing $N$ and then obtaining the chemical potential, we first set $n_0$ to a typical experimental value of the density. Then, using $\mu_0=gn_0$, we compute the number of particles $N$ by integrating the resultant GP wave function. The computed number of particles satisfies $N/L=n_0\left[1+O\left(\xi_0/L\right)+O\left(d/L\right)\right]$.

\subsection{Realistic optical lattice potential}

In order to simulate a more realistic scenario, we introduce the following two potentials:
on the left side, a Gaussian barrier centered at $x=0$ of the form $V_{L}(x)=U\exp(-2x^2/w_L^2)$, with $w_L=2~\mu\text{m}$, and on the right side, a realistic optical lattice centered at $x=L$, which has the form $V(x)=V_0\cos^{2}\left[k_L(x-L)\right]\exp\left[-2(x-L)^2/\tilde{w}^2\right]$.

We take the amplitudes of the confining potentials much larger than the chemical potential. We also take $L\gg\tilde{w}\gg w_L$, so that they are well separated in space and there is a large region where the potential is negligible and where we expect some kind of flat wave function. The width $w_L$ does not play a significant role in our simulations; we choose $w_L=2~\mu\text{m}$. In this way, we can set as boundary conditions for the numerical computation $A(0)=0$ and $A(L_{\rm{bc}})=0$, with $L_{\rm{bc}}$ sufficiently deep in the region where $V_0\exp[-2(x-L)^2/\tilde{w}^2]\geq\mu_{0}$. Once we have fixed the potential and the boundary conditions for the numerical calculation, we repeat the same process of the previous subsection by fixing $n_0$ to a typical experimental value and using the resulting value of $\mu_0$ to compute the number of particles $N$.

\section{Flowing condensate in a nonlinear optical lattice}

\label{app:nonlinearol}

The results of this section are partially based on Ref. \onlinecite{Wu2003}. The time-independent GP equation in an ideal {\it infinite} optical lattice whose potential has the same form of the long-time potential of Eq. (\ref{eq:TDPotential}), $V(x)= V_{\infty} \cos^2(k_L x )$, reads, after rescaling the wave function and the coordinates, $\Psi_0(z)\equiv \Psi(x)/\sqrt{n_r}$ (with $z\equiv 2k_L x$),
\begin{eqnarray}\label{eq:dimensionlessOLGPequation}
-\frac{1}{2}\frac{\partial^{2}\Psi_{0}}{\partial z^{2}}+v\cos(z)\Psi_{0}+c^{2}|\Psi_{0}|^{2}\Psi_{0} & = & \alpha \Psi_{0}\nonumber \\
%v=\frac{V_{0}}{8\hbar^{2}k_L^{2}/m},~c^{2}=
%\frac{gn_{r}}{4\hbar^{2}k_L^{2}/m},~ \alpha =\frac{\mu-\frac{V_{0}}{2}}{4\hbar^{2}k_L^{2}/m}\, ,
v=\frac{V_{\infty}}{2E_L},~c^{2}=
\frac{gn_{r}}{E_L},~ \alpha =\frac{\mu-V_{\infty}/2}{E_L}\, ,
\end{eqnarray}
where $n_r$ is the average atomic density, $\mu$ is the chemical potential, and $E_L=4\hbar^2k_L^2/m=8E_R$.
% is given after Eq. (\ref{eq:top-bottom}).

We look for solutions of the Bloch form
\begin{equation}
\Psi_{0}(z)=e^{iqz}y_{q}(z),
\end{equation}
with $y_{q}(z+2\pi)=y_{q}(z)$ periodic, because the non-linear term is periodic
for a Bloch-wave type solution. The normalization condition reads
\begin{equation}
\frac{1}{2\pi}\int_{0}^{2\pi}\mathrm{d}z~|\Psi_{0}(z)|^{2}
%=\frac{1}{2\pi}\int_{0}^{2\pi}
%\mathrm{d}z~|y_{q}(z)|^{2}
=1.\label{eq:olnormalization}
\end{equation}

The Brillouin zone is placed in the region
 $-1/2 < q < 1/2$. The equation for $y_{q}$ is:
\begin{equation}\label{eq:dimensionlessOLGPBlochequation}
-\frac{1}{2}\left(\frac{\partial}{\partial z}+iq\right)^{2}y_{q}+v\cos(z)y_{q}+c^{2}|y_{q}|^{2}y_{q}=\alpha_qy_{q},
\end{equation}
where we have allowed for a $q$-dependence of $\alpha$ defined in (\ref{eq:dimensionlessOLGPequation}). The linear (Schr\"odinger) regime is obtained when $c=0$. For $c^{2}>v$,
some extra non-linear Bloch waves appear. This generates a loop structure in the conduction band. In the systems analyzed in the present work,
$c^{2}\sim10^{-3}-10^{-4}$ and $v\sim10^{-1}$, hence $v\gg c^{2}$. As a consequence: (a) loops do not appear; (b) the system is close to the linear Schr\"odinger regime.

To compute the Bloch energy eigenvalues, we follow the method developed
in Ref. \onlinecite{Wu2003}. First, we perform a finite Fourier expansion of the periodic
function $y_{q}(z)$ of the form:
\begin{equation}\label{eq:Fouriercut}
y_{q}(z)=\sum_{n=-M}^{M}c_{n}e^{inz},
\end{equation}
where $M$ is a numerically enforced cut-off. After substitution of this solution in (\ref{eq:dimensionlessOLGPBlochequation})
and in (\ref{eq:olnormalization}), we get $2M+2$ equations for
$2M+2$ variables (the $2M+1$ values of the Fourier coefficients $c_{n}$ plus the eigenvalue $\alpha$). Instead of directly solving these non-linear
equations, it is more efficient to minimize the quadratic sum of the $2M+2$ equations,
\begin{equation}
S=\sum_{j=1}^{2M+2}f_{j}^{2}\label{eq:NonLinearOLQuadraticSum},
\end{equation}
where $f_{j}(c_{n},\alpha)=0$ (with $j=1,2\ldots2M+2$) are the equations to be solved. It is easy to see that all these equations are
real, so the coefficients $c_{n}$ can be chosen as real numbers and there is no need to use complex conjugates in (\ref{eq:NonLinearOLQuadraticSum}).

A given Bloch solution can be unstable, either dynamically or in the sense of Landau, as explained below.
The GP wave function is an extreme of the grand canonical Hamiltonian

\begin{widetext}
\begin{equation}
K\left[\Psi(z),\alpha\right]=\int\mathrm{d}z~ \left[ %-\Psi^{*}\frac{1}{2}\frac{\partial^{2}\Psi}{\partial z^{2}}
\frac{1}{2} \left| \frac{\partial \Psi}{\partial z}  \right|^2
+v\cos(z)|\Psi|^{2}+\frac{c^{2}}{2}|\Psi|^{4}-\alpha|\Psi|^{2} \right] \label{eq:olgrandcanonicalhamiltonian} \, .
\end{equation}
\end{widetext}

A superflow through the lattice is obtained when the actual solution $\Psi_0(z)$
minimizes the functional $K\left[\Psi(z),\alpha\right]$. When this is not the case, the system can
minimize its energy by the emission of excitations (phonons). The mean field
solutions of this last type are said to exhibit Landau instabilities.
These instabilities can be sought by expansion of $K\left[\Psi(z),\alpha\right]$
around $\Psi_0(z)$, i.e., $\Psi(z)=\Psi_{0}(z)+\delta\Psi(z)$.
The first-order term is automatically zero because $\Psi_{0}(z)$ solves the GP equation.
The quadratic terms reads
\begin{eqnarray}
\delta K & = & \frac{1}{2}\int\mathrm{d}z~[\delta\Psi^{*}~\delta\Psi]\Lambda\left[\begin{array}{c}
\delta\Psi\\
\delta\Psi^{*}
\end{array}\right]\nonumber \\
\Lambda & = & \left[\begin{array}{cc}
H' & L\\
L^{*} & H'
\end{array}\right]\nonumber \\
H' & = & -\frac{1}{2}\frac{\partial^{2}}{\partial z^{2}}+v\cos(z)+2c^{2}|\Psi_{0}|^{2}-\alpha_q\nonumber \\
L & = & c^{2}\Psi_{0}^{2} \, .
\end{eqnarray}
Landau instabilities correspond to negative eigenvalues of the Hermitian operator $\Lambda$. The corresponding eigenvalue equation is
\begin{equation}\label{eq:landaueig}
\Lambda\left[\begin{array}{c}
u\\
v
\end{array}\right]=\lambda\left[\begin{array}{c}
u\\
v
\end{array}\right] \, .
\end{equation}
By absorbing the exponential plane-wave factor of
the Bloch-type GP solution, $u(z)=e^{iqz}u_{q}(z)$
and $v(z)=e^{-iqz}v_{q}(z)$, we arrive at a new matrix
operator $\Lambda_{q}$ which is periodic. Applying Bloch's theorem in the form $u_{q}(z)=e^{ikz}u_{q,k}(z)$ and $v_{q}(z)=e^{ikz}v_{q,k}(z)$
with $u_{q,k}(z)$ and $v_{q,k}(z)$ periodic in $[0,2\pi]$, the
final eigenvalue equation reads
\begin{eqnarray}\label{eq:landaublocheig}
\Lambda_{q,k}\left[\begin{array}{c}
u_{q,k}\\
v_{q,k}
\end{array}\right] & = & \lambda_{q,k}\left[\begin{array}{c}
u_{q,k}\\
v_{q,k}
\end{array}\right]\nonumber \\
\Lambda_{q,k} & = & \left[\begin{array}{cc}
H^{''}_{k+q} & L_{q}\\
L_{q}^{*} & H^{''}_{k-q}
\end{array}\right]\nonumber \\
H^{''}_{k} & = & -\frac{1}{2}\left(\frac{\partial}{\partial z}+ik\right)^{2}+v\cos(z)+2c^{2}|y_{q}|^{2}-\alpha\nonumber \\
L_{q} & = & c^{2}y_{q}^{2} \, .
\end{eqnarray}

%For a given solution $\Psi_0(z)=e^{iqz}y_{q}(z)$, Landau instabilities appear when one of the eigenvalues $\lambda_{q,k}$ is negative.

Dynamical instabilities correspond to modes that grow exponentially with time. They are computed by looking for non-real eigenvalues of the BdG equations, which are formally similar to Eq. (\ref{eq:landaueig}):
\begin{equation}
M\left[\begin{array}{c}
u\\
v
\end{array}\right]=\epsilon\left[\begin{array}{c}
u\\
v
\end{array}\right] \, ,
\end{equation}
with $M=\sigma_z \Lambda$ (here $\sigma_z=\text{diag}(1,-1)$ is the usual Pauli matrix). Bloch's theorem also applies here and after a computation similar to that which has led to Eq. (\ref{eq:landaublocheig}), the eigenvalue equation reads
\begin{equation}\label{eq:bdgblocheig}
M_{q,k}\left[\begin{array}{c}
u_{q,k}\\
v_{q,k}
\end{array}\right] = \epsilon_{q,k}\left[\begin{array}{c}
u_{q,k}\\
v_{q,k}
\end{array}\right],
\end{equation}
with $M_{q,k}=\sigma_z \Lambda_{q,k}$. In addition to the dynamical stability analysis, the real eigenvalues of this operator can be used to compute the speed of sound in the optical lattice. When $q=0$, it can be proven that the the small wave-vector $k$ eigenvalues goes like $\epsilon=\pm s|k|$, with $s$ the speed of sound (here, in units of $2\hbar k_L/m$). On the other hand, the form of the Bloch-type solution of the GP equation as a function of both $q$ and $n_r$ can be directly used to compute the sound speed without the need to solve for the BdG equations \cite{Pethick2008}. Restoring dimensions by introducing $Q=2k_Lq$, it can be proven that
\begin{equation}\label{eq:OLsound}
s=\frac{\sqrt{\partial^2_n \mathcal{E} \partial^2_{Q} \mathcal{E}}}{\hbar}
\end{equation}
where $\partial_n$ denotes derivative with respect the mean density, $n_r$, $\partial_{Q}$ is the derivative with respect the pseudomomentum $Q$, with both derivatives evaluated at $Q=0$, and $\mathcal{E}$ is an average energy density given by
\begin{widetext}
\begin{eqnarray}\label{eq:energydensity}
\mathcal{E}&=&\frac{n_r}{d}\int_0^{d}\mathrm{d}x~y^*_q(2k_L x)\left[-\frac{\hbar^{2}}{2m}\left(\frac{\partial}{\partial x}+iQ\right)^{2} + V_{\infty}\cos^2(k_L x) +\frac{gn_r}{2}|y_q(2k_L x)|^2 \right] y_{q}(2k_L x) \nonumber \\
&=&n_r\mu-\frac{gn^2_r}{4\pi}\int_0^{2\pi}\mathrm{d}z~|y_q(z)|^4.
\end{eqnarray}
\end{widetext}
We will make use of this expression in Appendix \ref{app:olperturbationtheory}.

\section{Perturbative treatment of the nonlinearity in the optical lattice}

\label{app:olperturbationtheory}

In this Appendix, some of the results of Appendix \ref{app:nonlinearol} are perturbatively explored further.  In Eq. (\ref{eq:dimensionlessOLGPequation}) there are two dimensionless parameters, $v$ and $c^2$. The former is essentially the amplitude of the potential ($v\gg1$ is the tight-binding regime while $v\ll1$ corresponds the nearly-free-particle regime) and the latter is a measure of the strength of the interaction or nonlinearity. In the cases studied in this paper, $v\ll 1$. However, the forthcoming discussion applies to arbitrary values of $v$. The quantity
$\delta \equiv c^2 \ll 1$ is the small parameter of our perturbation theory. We expand in powers of $\delta$ both the Bloch wave $y_q(z)$ and the displaced and dimensionless chemical potential $\alpha_q$, which solves Eqs.
(\ref{eq:olnormalization})-(\ref{eq:dimensionlessOLGPBlochequation}) and is defined in Eq. (\ref{eq:dimensionlessOLGPequation}).  We obtain
\begin{eqnarray}
y_q(z) & = & \sum_{m=0}^\infty \delta^m y_{q}^{(m)}(z) \, , \nonumber \\
\alpha_q & = & \sum_{m=0}^\infty \delta^m \alpha_q^{(m)}\, , \label{eq:alpha-q}
\end{eqnarray}
which transforms Eq. (\ref{eq:dimensionlessOLGPBlochequation}) into [hereafter we omit the explicit $z$-dependence in $y_{q}(z), y_{q}^{(m)}(z)$]
\begin{eqnarray}
&& H^{(0)}_q y_{q}+\delta|y_{q}|^{2}y_{q} = \alpha_qy_{q} \nonumber \\
&& H^{(0)}_q\equiv  -\frac{1}{2}\left(\frac{\partial}{\partial z}+iq\right)^{2}+v\cos(z) \, .
\end{eqnarray}
In what follows, we focus on the lowest Bloch band and keep terms up to $O(\delta^{2})$.
The lowest-order term solves the linear Schr\"odinger equation, $H^{(0)}_qy_{q}^{(0)}=\alpha_q^{(0)}y_{q}^{(0)}$. Therefore, $y_{q}^{(0)}=\phi_{q,0}$ and $\alpha_q^{(0)}=\varepsilon_{q,0}$, where $\phi_{q,0}(z)$
is the corresponding eigenfunction for the lowest band, which involves Mathieu functions, and $\varepsilon_{q,0}$ its eigenvalue (note the use of the index $0$ for two different purposes: perturbative order is indicated in the superindex, while the band index comes in the subindex). We normalize $\phi_{q,0}$ according to (\ref{eq:olnormalization}).

The first-order corrections must satisfy
\begin{equation}
H^{(0)}_qy_{q}^{(1)}+|y_{q}^{(0)}|^2y_{q}^{(0)}=\alpha_q^{(0)}y_{q}^{(1)}+\alpha_q^{(1)}y_{q}^{(0)},
\end{equation}
which, using standard perturbation techniques, leads to
\begin{eqnarray}
\label{eq:perturbativeSolution}
\alpha_q^{(1)} & = & \frac{1}{2\pi}\int_0^{2\pi}\mathrm{d}z|\phi_{q,0}|^4 \nonumber \\
y_{q}^{(1)} & = & \sum_{n=1}^{\infty} \beta_n \phi_{q,n} \nonumber \\
\beta_n & = & \frac{\frac{1}{2\pi}\int_0^{2\pi}\mathrm{d}z~\phi^*_{q,n}|\phi_{q,0}|^2\phi_{q,0}}{\varepsilon_{q,0}-\varepsilon_{q,n}},
\end{eqnarray}
where $\{\phi_{q,n}\}_{n=1}^\infty$ are the Schr\"odinger eigenvectors of the rest of bands and $\varepsilon_{q,n}$ its corresponding eigenvalues, for a given value of $q$.

The second-order equation reads
\begin{widetext}
\begin{equation}
H^{(0)}_qy_{q}^{(2)}+2|y_{q}^{(0)}|^2y_{q}^{(1)}+y_{q}^{(0)2}y^{(1)*}_{q}=\alpha _q^{(0)}y_{q}^{(2)}+\alpha _q^{(1)}y_{q}^{(1)}+\alpha _q^{(2)}y_{q}^{(0)}.
\end{equation}
\end{widetext}
We note that $y_{q}^{(2)}$ is not needed to compute $\alpha_q^{(2)}$. Specifically, we find
\begin{eqnarray}
\label{eq:perturbativeSolutionW2}
\alpha _q^{(2)}&=&\frac{1}{2\pi}\int_0^{2\pi}\mathrm{d}z|\phi_{q,0}|^{2}(2\phi^*_{q,0}y_{q}^{(1)}+\phi_{q,0}y_{q}^{(1)*})\nonumber \\
&=&-3\sum_{n=1}^{\infty}|\beta_n|^2(\varepsilon_{q,n}-\varepsilon_{q,0}),
\end{eqnarray}
which is always negative.

Instead of invoking Mathieu functions, the numerical computation of the formulae presented in this Appendix  [Eqs. (\ref{eq:perturbativeSolution}), (\ref{eq:perturbativeSolutionW2})] can be easily performed in a Fourier representation.
As $y_q, \phi_{q,n}$ are periodic functions in $[0,2\pi]$, their Fourier expansion read $y_q(z)=\sum_{m=-\infty}^{\infty}c_me^{imz}$ and $\phi_{q,n}(z) = \sum_{m=-\infty}^{\infty}a_{n,m}e^{imz} $. Both the $c_m$ and the $a_{n,m}$ coefficients can be chosen real (see Appendix \ref{app:nonlinearol}). In this Fourier representation, $H^{(0)}_q$ is a tridiagonal matrix with elements $\mathbf{H}_{m,m\pm 1}=v/2$ and $\mathbf{H}_{m,m}=(m+q)^2 /2$. Multiplication by $|\phi_{q,0}|^{2}$ is represented by the matrix $\mathbf{r}_{m,p}=\sum_l a_{0,l}a_{0,l+m-p}$. The perturbative expansion of the Fourier components of solution to the GP equation reads $c_m=\sum_{n=0}^\infty \delta^n c_m^{(n)}$. In this Fourier basis, Mathieu's equation for all the bands is written as an eigenvalue-eigenvector matrix equation
\begin{equation}
\mathbf{H} \mathbf{a}_n  =  \varepsilon_{q,n} \mathbf{a}_n,
\end{equation}
where matrix multiplication is understood. The other previous results, Eqs. (\ref{eq:perturbativeSolution}), (\ref{eq:perturbativeSolutionW2}) can be written as:
\begin{eqnarray} \label{eq:fourier-pert}
\alpha_q^{(1)}&=& \mathbf{a}^{\intercal}_0\mathbf{r}\mathbf{a}_0\nonumber \\
\mathbf{c}^{(1)} & = & \sum_{n=1}^{\infty} \beta_n \mathbf{a}_n \nonumber \\
\beta_n & = & \frac{\mathbf{a}^{\intercal}_n\mathbf{r}\mathbf{a}_0}{\varepsilon_{q,0}-\varepsilon_{q,n}} \nonumber \\
\alpha_q^{(2)}&=&3\mathbf{a}^{\intercal}_0\mathbf{r}\mathbf{c}^{(1)}=-3\sum_{n=1}^{\infty}|\beta_n|^2(\varepsilon_{q,n}-\varepsilon_{q,0}).
\end{eqnarray}

Now we can use the perturbative results (\ref{eq:fourier-pert}) to give approximate closed expressions for some parameters of the optical lattice. To first order in $\delta$, the energy density is given by Eq. (\ref{eq:energydensity})
\begin{equation} \label{eq:E-1st-order}
\mathcal{E}
\simeq n_r\left(\mu-\frac{gn_r}{2}\alpha_q^{(1)}\right).
\end{equation}
We can write
%$\mu=E_{\rm lat}\alpha+V_0/2\simeq E_{\rm lat}\alpha_q^{(0)}+V_0/2+E_{\rm lat}\alpha_q^{(1)}\delta=\mu^{(0)}+E_{\rm lat}\alpha_q^{(1)}\delta$
\begin{eqnarray}\label{eq:oldmu}
\mu & = & E_L\alpha+\frac{V_{\infty}}{2}\simeq E_L\alpha_q^{(0)}+\frac{V_{\infty}}{2}
+E_L\alpha_q^{(1)}\delta \nonumber \\
& = & \mu^{(0)}+E_{L}\alpha_q^{(1)}\delta \, ,
\end{eqnarray}
where $E_L$ is defined after Eq. (\ref{eq:dimensionlessOLGPequation}).
On the other hand, for the non-interacting chemical potential we have (assuming $q\ll1/2$):
\begin{equation}
\mu^{(0)}=\mu^{(0)}(Q) \simeq E_{\rm{min}}+\frac{\hbar^{2}Q^2}{2m^*} \, ,
\end{equation}
where $E_{\rm{min}}$ is the bottom of the conduction band as defined in the main text, $m^*$ is the effective mass, and we recall $Q=2k_Lq$.
Thus we can rewrite Eq. (\ref{eq:oldmu}) as:
\begin{equation}\label{eq:olmu}
\mu=E_{\rm{min}}+\frac{\hbar^{2}Q^2}{2m^*}+gn_r\alpha_q^{(1)} \, .
\end{equation}
Using (\ref{eq:olmu}) we can rewrite (\ref{eq:E-1st-order}) as
\begin{equation}
\mathcal{E}\simeq n_rE_{\rm{min}}+n_r\frac{\hbar^{2}Q^2}{2m^*}+\frac{gn^2_r}{2}\alpha_q^{(1)} \, .
\end{equation}
Computing the derivatives to lowest order in $\delta$, we arrive at:
\begin{eqnarray}
\frac{\partial^2 \mathcal{E} }{\partial Q^2} &\simeq& n_r\frac{\hbar^2}{m^*} \, , \nonumber \\
\frac{\partial^2 \mathcal{E}}{\partial n_r^2}&\simeq& g\alpha_{0}^{(1)} \, .
\end{eqnarray}
Now we compute the speed of sound using Eq. (\ref{eq:OLsound}) and obtain:
\begin{equation}\label{eq:PerturbativeSound}
s=\sqrt{\frac{gn_r}{m^*}\alpha_{0}^{(1)}}=\sqrt{\frac{gn_r}{m}}\sqrt{\frac{m}{m^*}\alpha_{0}^{(1)}} \, .
\end{equation}
Similar results appear in Ref. \onlinecite{Kramer2003} and references therein. The first square root is the speed of sound in the absence of the optical lattice. The second factor on the right takes into account the presence of the optical lattice and is practically unity for $v\ll 1$. Specifically, we can write:
\begin{eqnarray}
m^*&=&m\left(1+8v^2+O(v^4)\right)\nonumber \\
\alpha_q^{(1)}&=&1+\frac{8v^2}{(1-4q^2)^2}+O(v^4),
\label{m-alpha-1}
\end{eqnarray}
and then $\sqrt{m\alpha_{0}^{(1)}/m^*}=1+O(v^4)$, so $s\simeq\sqrt{gn_r/m}$, which is the usual expression for the speed of sound. Equation (\ref{eq:PerturbativeSound}) can also be interpreted as the sound velocity arising in a system with an effective constant coupling $g_{\rm{eff}}=g\alpha_{0}^{(1)}$ and effective mass $m^{*}$ \cite{Carusotto2002}.

The current is also conserved for a stationary solution of the GP equation and is given, to lowest order in $\delta$ and $Q$, by:
\begin{equation}\label{eq:olcurrent}
j=\frac{1}{\hbar}\frac{\partial\mathcal{E}}{\partial Q}=n_r\frac{\hbar Q}{m^*} \, .
\end{equation}
In the nearly-free atom approximation, where the relative oscillations of the density around the mean value are small, we can write $j\simeq n_r\bar{v}$ were $\bar{v}$ is a locally averaged flow velocity (not to be confused with the dimensionless parameter $v$). Then, we have $\bar{v}\simeq\hbar Q/m^*$ and we can rewrite Eq. (\ref{eq:olmu}) in a more appealing form:
\begin{equation}\label{eq:olhdmu}
\mu=E_{\rm{min}}+\frac{1}{2}m^*\bar{v}^2+m^*s^2 \, .
\end{equation}
The physical interpretation of this equation is straightforward: the chemical potential in the optical lattice is the sum of the energy of the bottom of the conduction band plus the contribution of the kinetic energy and the interaction energy, both with $m^*$ instead of $m$.

As explained at the end of Appendix \ref{app:nonlinearol}, the same result for the speed of sound can be obtained by solving the BdG equations (\ref{eq:bdgblocheig}) perturbatively to first order in $\delta$. When $q=0$ (which implies that the GP solution $y_0(z)$ can be taken as real), we perform an expansion of the spinors in terms of solutions to the Schr\"odinger equation,
\begin{equation}
\left[\begin{array}{c}
u_{0,k}(z)\\
v_{0,k}(z)
\end{array}\right] =\sum_{n=0}^\infty \phi_{k,n}(z)\chi_{k,n},
\end{equation}
where $\chi_{k,n}$ are spinors of constant ($z$-independent) coefficients. The matrix operator $M_{0,k}$
introduced in (\ref{eq:bdgblocheig}) can be written, to first order in $\delta$, as $M_{0,k}=M^{(0)}_{k}+ M^{(1)}\delta$  with:
\begin{eqnarray}
M^{(0)}_{k} & = & \left[\begin{array}{cc}
H^{(0)}_{k}-\varepsilon_{0,0} & 0\\
0 & -H^{(0)}_{k}-\varepsilon_{0,0}
\end{array}\right] \\
M^{(1)} & = & \left[\begin{array}{cc}
2\phi^2_{0,0}(z)-\alpha_{0}^{(1)} & \phi^2_{0,0}(z)\\
-\phi^2_{0,0}(z) & -2\phi^2_{0,0}(z)+\alpha_{0}^{(1)}
\end{array}\right] \, , \nonumber
\end{eqnarray}
$\phi_{0,0}(z)$ being the Schr\"odinger solution for the bottom of the lowest band. Note that $\phi_{k,n}(z)$ are eigenfunctions of $H^{(0)}_{k}$. A matrix equation for the perturbative expansion of the $\chi_{k,0}$ spinors to first order can be obtained by projecting onto the lowest Bloch eigenfunction, $\phi^*_{k,0}(z)$:
\begin{widetext}
\begin{eqnarray}\label{eq:effectiveolbdg}
\epsilon(q=0,k) \chi_{k,0} & = & \left[\begin{array}{cc}
\varepsilon_{k,0}-\varepsilon_{0,0}+(2J(k)-\alpha^{(1)}_{0})\delta & J(k) \delta \\
-J(k) \delta & -\varepsilon_{k,0}+\varepsilon_{0,0}-(2J(k)-\alpha^{(1)}_{0}) \delta
\end{array}\right] \chi_{k,0} \nonumber \\
J(k) & = & \frac{1}{2\pi}\int_0^{2\pi}\mathrm{d}z|\phi_{k,0}|^2|\phi_{0,0}|^2.
\end{eqnarray}
\end{widetext}
Restoring units for $k$ by using $K=2k_Lk$, expanding to lowest order near $K=0$, and neglecting corrections $O(\delta)$ to $m^{*}$, the eigenvalues are approximated as
\begin{equation}
%\epsilon(0,k) = \pm\sqrt{\left(\frac{\hbar^2k^2}{2m^*}\right)^2+gn_r\alpha_{0}^{(1)}\frac{\hbar^2k^2}{m^*}}
\epsilon(0,K) = \pm\left[
\left(\frac{\hbar^2K^2}{2m^*}\right)^2+gn_r\alpha_{0}^{(1)}\frac{\hbar^2K^2}{m^*}
\right]^{\frac{1}{2}} \, ,
\end{equation}
which for small $K$ gives $\epsilon(0,K)\simeq \hbar sK$ with $s$ given by (\ref{eq:PerturbativeSound}).

The previous results can be used to compute the corrections to the width of the lowest band, which by using (\ref{eq:effectiveolbdg}) leads to
%(units are restored)
\begin{equation}
\Delta^{\rm BdG}_c \simeq \Delta_{c}+\left[2J(1/2)-\alpha_{0}^{(1)}\right] gn_r \, ,
\end{equation}
where $\Delta_{c}$ is the Schr\"odinger bandwidth and
we have used $\Delta_{c}\gg gn_{r}$, which is true in all the situations considered in the present work.

\section{Numerical methods: Crank-Nicolson method and absorbing boundary conditions} \label{app:crnicabc}

The numerical computation of the time evolution of the system has been made using the Crank-Nicolson method, as in Ref. \onlinecite{PhysRevA.76.063605}.
The spatial interval $[0,L_{g}]$ is divided into $N+2$ equally spaced points separated by a distance
$\Delta x=L_{g}/(N+1)$, and the time interval $[0,t]$ into steps of size $\Delta t$. Hence, we
write the grid points as
\begin{eqnarray}
x_{j} & = & j\Delta x~j=0,1\ldots N+1\nonumber \, \\
t_{k} & = & k\Delta t~k=0,1\ldots n \, .
\end{eqnarray}
Here we use units such that $\hbar=m=\xi_0=1$, and rescale the wave function by extracting the factor $\sqrt{n_0}$. The GP equation can be then written as
\begin{eqnarray}\label{eq:numericalGP}
i\frac{\partial\Psi (x,t)}{\partial t}& = & H(x,t)\Psi(x,t) \\
H(x,t) & = & -\frac{1}{2}\frac{\partial^{2}}{\partial x^{2}}+V(x,t)+|\Psi(x,t)|^{2}-1 \nonumber
\end{eqnarray}
where the "$-1$" comes from subtracting the initial chemical potential and $V(x,t)$ is the time-dependent potential.
Here, $H(x,t)$ plays the role of an effective Hamiltonian. If we define the
spatial vector with the discretized values of the wave function in a given time $t_{k}$ as $\mathbf{\Psi}_{k}$, with
components $\Psi_{k}^{j}=\Psi(x_{j},t_{k})$, and using
\begin{widetext}
\begin{eqnarray}
\Psi\left(x,t+\frac{\Delta t}{2}\right) & = & \frac{\Psi\left(x,t+\Delta t\right)+\Psi(x,t)}{2}+O\left(\Delta t^{2}\right)\nonumber \\
\frac{\partial\Psi}{\partial t}\left(x,t+\frac{\Delta t}{2}\right) & = & \frac{\Psi\left(x,t+\Delta t\right)-\Psi(x,t)}{\Delta t}+O\left(\Delta t^{2}\right)\nonumber \\
\frac{\partial^{2}\Psi}{\partial x^{2}}\left(x,t\right) & = & \frac{\Psi\left(x+\Delta x,t\right)+\Psi\left(x-\Delta x,t\right)-2\Psi(x,t)}{\Delta x^{2}}+O\left(\Delta x^{2}\right),
\end{eqnarray}
we can write, up to second order in $\Delta x$ and $\Delta t$, a
discrete version of (\ref{eq:numericalGP})
\begin{eqnarray}\label{eq:discreteGP}
i\frac{\mathbf{\Psi}_{k+1}-\mathbf{\Psi}_{k}}{\Delta t} & = & \mathbf{H}_{k+\frac{1}{2}}\frac{\mathbf{\Psi}_{k+1}+\mathbf{\Psi}_{k}}{2}\nonumber \\
\left(\mathbf{H}_{k+\frac{1}{2}}\mathbf{\Psi}\right)^{j} & = & -\frac{\Psi^{j+1}+\Psi^{j-1}-2\Psi^{j}}{2(\Delta x)^{2}}+V_{k+\frac{1}{2}}^{j}\Psi^{j}+|\Psi_{k+\frac{1}{2}}^{j}|^{2}\Psi^{j}-\Psi^{j}\nonumber \\
V_{k+\frac{1}{2}}^{j} & = & V\left(x_{j},t_{k}+\frac{\Delta t}{2}\right) \, .
\end{eqnarray}
\end{widetext}

This can be written in matrix form
\begin{eqnarray}
\mathbf{M}_{2}\mathbf{\Psi}_{k+1} & = & \mathbf{M}_{1}\mathbf{\Psi}_{k}\nonumber \\
\mathbf{M}_{1,2} & = & 1\mp i\mathbf{H}_{k+\frac{1}{2}}\frac{\Delta t}{2} \, ,
\end{eqnarray}
where
\begin{equation}\label{eq:mmatrices}
\mathbf{M}_{1,2}=\left[\begin{array}{cccccc}
\ddots & \ddots & \ddots\\
 & \pm A & 1\mp B_{k}^{j} & \pm A\\
 &  & \pm A & 1\mp B_{k}^{j+1} & \pm A\\
 &  &  & \ddots & \ddots & \ddots
\end{array}\right] \, ,
\end{equation}
with
\begin{equation}
A=\frac{i\Delta t}{4\Delta x^{2}},~B_{k}^{j}=i\frac{\Delta t}{2}\left(\frac{1}{\Delta x^{2}}+V_{k+\frac{1}{2}}^{j}+|\Psi_{k+\frac{1}{2}}^{j}|^{2}-1\right) \, .
\end{equation}
Because we ignore the value of $\Psi_{k+\frac{1}{2}}^{j}$ in the non-linear term,
we use a corrector-predictor method, which consists in performing an additional
step in every time iteration. In the first iteration, we use $\Psi_{k}^{j}$
instead of $\Psi_{k+\frac{1}{2}}^{j}$ in order to obtain
a value $\bar{\Psi}_{k+1}^{j}$. Next, we perform a new iteration
taking $\Psi_{k+\frac{1}{2}}^{j}=\left(\bar{\Psi}_{k+1}^{j}+\Psi_{k}^{j}\right)/2$
to obtain the final value $\Psi_{k+1}^{j}$.

The main advantage of this integration scheme is that the obtention of $\Psi_{k+1}^{j}$
only requires the resolution of a tridiagonal system of equations, which is computationally very efficient
(the number of operations grows like $N$).

The hard-wall boundary conditions reads
\begin{equation}\label{eq:hardwall}
\Psi_{k}^{l}=0,\,\,~l=0,N+1,
\end{equation}
and this can be easily implemented by suppressing the first and the last
columns of the M matrices in (\ref{eq:mmatrices})
\begin{equation}\label{eq:hardwallcrnic}
\mathbf{M}_{1,2}=\left[\begin{array}{ccc}
1\mp B_{k}^{1} & A\\
 & \ddots & \ddots\\
 & A & 1\mp B_{k}^{N}
\end{array}\right]
\end{equation}

On the other hand, any boundary condition imposed at the final point of the grid ($x=L_{g}$)
will induce reflections which are unwanted because our goal is to simulate a semi-infinite supersonic region. To minimize those spurious reflections, one can use complex absorbing potential (CAP)
at the grid boundaries \cite{Muga2004357}. Instead of that, we make use of the alternative, so-called
ABC (Absorbing Boundary Conditions) \cite{PhysRevE.74.037704,PhysRevE.78.026709}. This method is based on the linearization of the dispersion relation in the boundary in order to achieve the relation
corresponding to an outgoing plane wave. Both ABC and CAP are very useful because
they not only prevent the artificial reflection of the waves,
but also permit to reduce the size of the supersonic zone.

The point $x=L_{g}$ is placed in the supersonic zone, where there is no potential.
In addition, we expect that the non-linear term in (\ref{eq:numericalGP}) can be neglected. This
means that the effective Hamiltonian $H$ in this region is the usual free (Schr\"odinger)
Hamiltonian and Eq. (\ref{eq:numericalGP}) can be written as
\begin{equation}
i\frac{\partial\Psi}{\partial t}=-\frac{1}{2}\frac{\partial^{2}\Psi}{\partial x^{2}}-\Psi\label{eq:spGP}
\end{equation}
which implies the dispersion relation
\begin{equation}
\omega=\frac{k^{2}}{2}-1
\end{equation}
On the supersonic side and in the quasi-stationary regime, the wave
function is well peaked in momentum space around a value $k_{0}\simeq\sqrt{2E_{\rm min}}$, where $E_{\rm min}$ is the energy of the bottom of the first
conduction band. By linearizing the dispersion relation around $k_{0}$ and expressing this relation
in terms of derivatives, one can get
\begin{equation}\label{eq:eqABC}
i\frac{\partial\Psi}{\partial t}=-ik_{0}\frac{\partial\Psi}{\partial x}-
\left(\frac{k_{0}^{2}}{2}+1\right)\Psi \, .
\end{equation}
We replace the hard-wall boundary condition at $j=N+1$, Eq. (\ref{eq:hardwall}), by the discrete version of Eq. (\ref{eq:eqABC}) at $j=N$, hence there are $N+1$ variables ($\Psi_{k}^{j}, j=1\ldots N+1$) and $N+1$ equations, corresponding to $N$ equations resulting from Eq. (\ref{eq:discreteGP}) for $j=1\ldots N$ and the ABC equation. We can regard the point $x_{N+1}$ as a ghost point because the GP equation is not properly defined there and the ABC (\ref{eq:eqABC}) is the corresponding equation for this point \cite{PhysRevE.78.026709}.
Following these considerations, we can easily implement the ABC condition by adding a new row to the matrices M, which now are of size $\left(N+1\right)\times\left(N+1\right)$
and of the form
\begin{equation}
\mathbf{M}_{1,2}=\left[\begin{array}{cccc}
\ddots & \ddots\\
 & \pm A & 1\mp B_{k}^{N} & \pm A\\
 & \pm C & 1\pm D & \mp C
\end{array}\right]
\end{equation}
with
\begin{equation}
C=\frac{k_{0}\Delta t}{4\Delta x},~D=i\frac{\Delta t}{2}\left(\frac{k_{0}^{2}}{2}+1\right)
\end{equation}

Due to the finite size of the grid and to the nonzero width of the momentum distribution in the supersonic region, the absorption is not perfect. We have found however that, in practice, the small spurious reflections have no effect on the final results.

\bibliographystyle{apsrev}
\bibliography{HawkingOLMFJR}

\end{document}